# MULTIGENERATIONAL EFFECTS OF SMALLPOX VACCINATION

Volha Lazuka* and Peter Sandholt Jensen**

**Abstract**

Using Swedish register data spanning 250 years, we estimate multigenerational effects of smallpox vaccination on longevity and occupational achievements. Employing mother fixed-effects, difference-in-differences, and shift-share instrumental-variables designs, we find vaccination improves outcomes for three generations. We explore mechanisms through which benefits transmit across generations, finding evidence consistent with both improved health behaviors and epigenetic inheritance. Effects persist even in milder disease environments, demonstrating vaccination's lasting benefits beyond epidemic contexts. These findings underscore the importance of accounting for multigenerational returns when evaluating early-life health interventions.

* Corresponding author: vola@sam.sdu.dk, University of Southern Denmark, Lund University, and IZA.

** Linnaeus University.

**Disclosure Statement**

The authors declare that they have no relevant or material financial interests that relate to the research described in this paper We are grateful to the participants of the EEA Annual Congress, ASSA Annual Meeting, PAA Annual Meeting, Economic Society of Population Economics Conference, and of the seminars at the UC Davis, London School of Economics, and the University of Southern Denmark. Volha acknowledges funding from the Marie-Curie Individual Fellowship (grant No 101025481).

**Data Availability Statement**

The data are confidential from Demographic Data Base and the Scanian Economic-Demographic Database, but others can access it. For data access, permission can be obtained from the Ethical Review Board (https://www.epn.se/en/start/). Then applications have to be made to the Centre for Economic Demography (Lund University) and the Centre for Demographic and Ageing Research (Umeå University). Program files are available for replication, and we are willing to assist with replication efforts to the best of our abilities.

## Introduction

The recent global disruptions caused by emerging infectious diseases underscore the critical importance of vaccines in protecting populations against these diseases. Live vaccines may also protect against unrelated diseases, known as "non-specific" vaccine effects, with childhood vaccination enhancing the immune system through epigenetic and metabolic changes (Benn et al. 2023; de Bree, L C J et al. 2018). Recent experimental evidence from primates demonstrates that the benefits of early-life interventions can exhibit intergenerational complementarity, where advantages are present only for offspring of mothers who themselves received the intervention (Dettmer et al. 2025). In parallel to epidemiological findings, economic studies have explored childhood vaccination's impact on adult labor market outcomes in a human capital framework (Barteska et al. 2023; Atwood 2022; Bütikofer and Salvanes 2020). The full effects of vaccination remains underestimated because they extend across lifetimes and generations, yet the related research is scarce. Understanding these multigenerational effects is essential for assessing vaccination benefits and informing public health decisions related to vaccination provision and withdrawal.

Establishing causal effects of vaccination across generations presents formidable empirical challenges. First, families who choose to vaccinate may differ systematically from those who do not in ways that independently affect outcomes--such as trust in authorities, socio-economic status, or health-seeking behaviors. These differences can bias estimates if not properly addressed. Selective attrition over long observation periods can introduce bias if migration patterns differ systematically between vaccinated and unvaccinated individuals.

Second, to credibly estimate intergenerational effects, a researcher must trace individuals and their descendants across multiple generations--requiring exceptionally detailed longitudinal data spanning centuries. Distinguishing the mechanisms of intergenerational transmission requires separating biological factors (such as epigenetic inheritance) from behavioral factors (such as the transmission of health-seeking behaviors). Any vaccination intervention typically occurs alongside other health and economic changes, making it difficult to isolate the specific impact of vaccination from concurrent developments.

In this study, we estimate the impact of smallpox vaccination on the longevity and economic well-being of three generations, spanning the period from 1790 to 2016. We leverage the smallpox vaccination campaign in Sweden as a quasi-experiment, employing mother fixed effects, difference-in-differences, and shift-share instrumental-variables strategies to establish causal effects. We find that smallpox vaccination in early childhood enhances both longevity and occupational achievements of Generation 1 as well as of their children (Generation 2) and grandchildren (Generation 3). We explore the mechanisms in causal mediation analysis and find that these benefits arise from both improved health behaviors and epigenetic factors.

Our analysis relies on three complementary identification strategies. First, we employ mother fixed effects, which compare biological siblings within the same family, one of whom received vaccination while the other did not. This effectively removes all fixed, unobservable family-related factors that may affect both treatment assignment and long-term outcomes. Second, we implement a difference-in-differences design that exploits two sources of variation: the varying exposure of different cohorts to the introduction of the smallpox vaccine in 1801, and the relatively larger benefits for individuals born in parishes with greater availability of church assistants and musicians--the primary vaccinators--compared to those born in parishes with lower availability.

Third, we adapt a shift-share instrumental-variables approach that captures the exogenous variation in children's smallpox vaccination. Our instrument interacts the parish-level number of church assistants and musicians with the national rate of change in their employment. This strategy leverages the fact that vaccination in Sweden was carried out by low-skilled clergy who otherwise did not perform public health duties, ensuring that their employment patterns satisfy the exclusion restriction. We apply this design for Generation 2 and 3. We address potential selective migration by adjusting estimates using Heckman's two-stage selection procedure.

Our data come from unique register-based datasets containing health, demographic, and socio-economic information on residents from 79 different parishes in Sweden, spanning from the 18th to the 21st centuries, including their descendants. We obtained data from parishes representing northern, central, and southern Sweden that are based on reconstructed life and family histories of parish residents, with variables at the individual level in the same metrics across cohorts.

The analytic sample for Generation 1 includes individuals born between 1790 and 1820. Generation 2 includes the children of Generation 1, born between 1805 and 1865, and Generation 3 includes the grandchildren of Generation 1, born between 1820 and 1910. In total, we tracked the full life cycles of three generations, which amounts to 141,067 individuals, up until their death, out-migration, or reaching the age of 100. Our key treatment variable is whether an individual belonging to Generation 1 was vaccinated against smallpox by the age of 2 years. We measure longevity as the count of years lived after age two, and we also examine disability-free years lived, occupational scores, and literacy.

To implement our empirical strategies, we construct an instrument based on the interaction between the number of church assistants and musicians at the parish level and the national rate of change in their employment. The historical context supports our identification: the state authorities did not allow local physicians to monopolize the vaccination process, fearing low uptake. As a solution, starting in 1805, low-skilled church personnel, commonly with no prior involvement in health or epidemic matters, were obliged to attain the skill of vaccination. Data from the 1810s indicate that over 60% of those administering vaccinations were church assistants or church musicians. This unique institutional feature ensures that variation in vaccinator availability provides quasi-random variation in vaccination uptake.

We address the key challenges to causal identification through several strategies. To handle unobservable family-level selection, we employ mother fixed effects that compare siblings within families. To address time-varying confounders at the parish level, we include parish fixed effects, region-by-cohort fixed effects, and interactions between cohort dummies and parish-level characteristics such as the number of midwives, smallpox death rates, urbanization measures, and economic indicators. We assess parallel pre-trends through event-study specifications and by examining leads of our instrument. To correct for selective migration, we implement Heckman's correction procedure using rye prices and month of birth as predictors of sample inclusion.

Our findings reveal substantial and persistent effects of smallpox vaccination across generations. Smallpox vaccination adds 11 years of life to the first generation and 2 and 1 years to the second and third generations. To put such results in perspective, vaccination in childhood in historical Sweden produces similar effects for longevity as quitting smoking in today's context. We establish explicit "non-specific" vaccination effects: while mortality from smallpox is reduced the most, there are negative effects on mortality from other causes. All three generations studied died primarily from causes other than smallpox, indicating that vaccination's protective effects extended beyond the targeted disease.

Vaccination also improves economic outcomes across generations. We find significant positive effects on occupational achievements, with these effects transmitted to subsequent generations at reduced magnitudes. Vaccination increases disability-free life expectancy by 11 years for Generation 1, 8 years for Generation 2, and 4 years for Generation 3. Causal mediation analysis reveals that more than half of the transmitted effects are attributed to nurture, as vaccinated individuals are more likely to vaccinate their children across generations, while epigenetic factors account for the remainder. These results are robust to numerous specification checks and consistent across our three identification strategies.

Our paper makes two key contributions. First, it contributes to the literature on intergenerational transmission of health. While several studies have attempted to derive the causal impacts of health transmission by relying on environmental shocks as sources of exogenous variation (East et al. 2023; Cook, Fletcher, and Forgues 2019), our knowledge of whether health shocks for one generation causally determine the outcomes of subsequent generations remains extremely scarce. A few recent studies have examined transgenerational effects of nutritional shocks on grandchildren (Maystadt and Migali 2021; Sari, Moilanen, and Sommerseth 2021), finding evidence of biological transmission across three generations. We contribute to this literature by tracing the effects of a positive health shock over the full life cycles of three generations. Moreover, we decompose these intergenerational effects into components driven by behavioral factors (health-seeking behaviors) and epigenetic factors, providing insights into the mechanisms underlying intergenerational health transmission.

Our paper also contributes to the literature on long-term effects of historical health interventions. While there exists an extensive body of literature on the long-term effects of health interventions (Anderson et al. 2020; Egedesø, Hansen, and Jensen 2020; Lazuka 2020), there is a smaller, yet steadily expanding, literature focused on vaccinations that occurred further back in history (Barteska et al. 2023; Atwood 2022; Bütikofer and Salvanes 2020). We contribute to this literature by examining the vaccination campaign, which is the world's first documented public health initiative. Our findings demonstrate that even in milder disease environments like those seen today, vaccination delivers lasting advantages, highlighting its long-term benefits beyond epidemic contexts.

# 1. The Context of Smallpox Vaccination in Sweden

## *1.1 Introduction of Smallpox Vaccination*

In 1798, Edward Jenner published his successful smallpox vaccination method, initially vaccinating a boy with cowpox then, after eight weeks, administering smallpox without adverse effects, confirming efficacy (Baxby 1985). Vaccination reached Sweden by December 7, 1801 (Riksarkivet 1802-1812). Swedish reformers considered it "among the greatest inventions ever, which—when it has increased in confidence—will be the supreme happiness of the human race and the triumph of medicine" (Hedin 1802). First Swedish vaccinations occurred in late 1801; from 1803, Stockholm's Inoculation House maintained fresh vaccine for nationwide efforts.

Pre-vaccination, inoculation—deliberate smallpox (not cowpox) skin infection—served as preventive measure. Though introduced in Britain in 1721, Swedish use began in 1756. Historical records suggest inoculation never gained wide acceptance due to death risk (Pettersson 1912). Country records and our data confirm low Swedish uptake: less than 0.01% of parishioners were inoculated between 1769–1800 (Riksarkivet 1769-1801).

Sweden's 1801 vaccination introduction had remarkable features exploited in our empirical design. First, vaccination focused on children, typically around age 2 (Riksarkivet 1802-1812). From March 1816, parents were required to vaccinate children under 2, with fines for non-compliance. Unable-to-pay parents faced imprisonment on water and bread diets.

Second, vaccination was nearly free. Vaccinators were not permitted to charge parents for vaccinating children. Local compensation included parish salaries, small wealthy parent fees, poor relief payments, or medals (Sköld 1996). No discernible social class differences existed in Swedish vaccination practice (Dribe and Nystedt 2003).

Finally, in 1804 every parish was instructed to appoint a vaccinator. State authorities prevented local physicians from monopolizing vaccination, fearing low uptake and voluntary fee charges. The vaccine could be safely administered by non-medical individuals—"anyone, without prior experience but with good interpersonal skills, common sense, and the ability to read and write, who would simply need to acquire a few skills" (Ekelund 1804). From 1805, low-skilled church personnel with no prior health/epidemic involvement were obliged to attain vaccination skills. 1810s data show over 60% of vaccinators were church assistants or musicians, followed by priests (12%), upper-class women (10%), and midwives, physicians, and others (each 5%) (Sköld 1996). Unlike local demographic and cultural factors, church personnel availability and employment emerges as the sole factor highly correlated with vaccination uptake.

### *1.2 Disease Environment and Vaccine Uptake*

Smallpox was the main childhood disease and death cause pre-vaccination. In Sweden, widespread *Variola major*, a highly contagious airborne type, affected children or persons lacking natural immunity. The historically unchanged virus caused two weeks of suffering (headache, fever, backache, vomiting, diarrhea), followed by pustules (Fenner et al. 1988). Case fatality rates reached 20% among infected persons and 55% among children below age 2 (Sköld 1996). Survivors often bore lifelong pockmarks and complications including blindness, baldness, limb deformities, infertility, and respiratory, gastrointestinal, and central nervous system conditions.

No effective treatment exists against smallpox due to viral resistance to temperatures below 60 degrees and nutrition independence, causing losses regardless of food access or family conditions. Pregnant women transmit infection, not protective antibodies, to fetuses, causing premature neonatal death. Children must develop their own immunity or receive vaccination for protection. Vaccination, including earlier iterations, offers approximately 95% effectiveness (Fenner et al. 1988).

Swedish smallpox mortality age patterns changed dramatically with 1801 vaccination introduction (Figure A1, Appendix A). Rates calculated from population counts for analyzed regions mirror national figures (Pettersson 1912). Despite significant unknown death causes, primary infectious disease symptoms were recognizable to registrars (priests, doctors); thus, smallpox death age patterns have high relative accuracy (Bengtsson and Lindstrom 2000). Between 1790–1800, smallpox mortality followed an L-shaped pattern: approximately 3% of children under three died (16% of all causes), with decreasing rates among older children and few adult deaths. This indicates societies where older individuals acquired natural immunity without passing it to children. Within decades post-1801, the age pattern flattened significantly; less than 0.2% of children died from smallpox.

Figure 1 shows cohort vaccination rates by age two for nineteenth-century Sweden, encompassing all studied generations. Vaccine uptake was gradual for decades post-1801, then stabilized at 85%. Mandatory vaccination in 1816 only modestly increased uptake among small children, suggesting most uptake associated with other factors—suggestively vaccinator numbers rather than mandatory law. No one died from smallpox in the 1890s, causing vaccination rates to decline. Mortality and vaccination patterns mirror those for Sweden overall (Sköld 1996).

[Figure 1 about here]

Why did not vaccination rates reach 100% despite mandatory status? Historical narrative suggests compulsory vaccination law was a threat making most parents comply (Pettersson 1912). Yet historical examples of executed fines are rare. Anti-vaccination opposition was lower in Sweden than other European countries, with the first known petition appearing half a century post-vaccination start. Nevertheless, some spread messages that smallpox was religious sin, and local authorities were reluctant to implement policy and conflict with religiously-motivated refusers (Sköld 1996). Another hesitance source was false stories about negative vaccine consequences spread by vagabonds and beggars. One local doctor classified non-vaccinating parents' cases as: laziness, pleasure from defying law, and vaccination consequence fears (Landsarkivet i Lund 1805-1827).

### *1.3 Non-Specific Vaccination Effects in the Historical Narrative*

Many contemporaries believed smallpox elimination would yield little gain as other diseases would replace it (Hofsten and Lundström 1976). Historical evidence from Sweden and elsewhere suggests the opposite—the vaccine combated both smallpox and other infectious diseases, known as "non-specific" vaccination health effects. Mayr (2004) cites circumstantial evidence from German and Austrian vaccinators reporting that "eye and ear disorders not only improved but also disappeared, and that chronic diseases vanished amongst the vaccinees." He notes vaccinated persons are less susceptible to infectious diseases (measles, scarlet fever, and whooping cough, even syphilis) than non-vaccinated persons. For Sweden, annual doctor reports show multiple indications of close associations between high vaccination rates and reduced infectious disease beyond smallpox (Riksarkivet 1796-1820). Moreover, lifetime gains for vaccinated children may emerge from improved disease environments. About 1% of smallpox survivors develop vivid lifelong complications (blindness, limb deformities, infertility, respiratory/gastrointestinal/central nervous system conditions) (Sköld 1996).

## 2. Data

### *2.1 Microdata for Three Generations.*

***Study area and sample.*** We investigate whether vaccinating Generation 1 against smallpox positively affects their lifelong well-being and that of Generations 2 and 3, using high-quality multi-generational data spanning extended time and age ranges.

Our data comprise unique register-based datasets with health, demographic, and socio-economic information from 79 Swedish parishes (18th–21st centuries). Sources include the Demographic Data Base for northern/central Sweden (CEDAR 2021; CEDAR 2022), and the Scanian Economic-Demographic Database for southern Sweden (Bengtsson et al. 2021). These datasets reconstruct life and family histories from high-quality archival records in geographically compact areas, minimizing regional biases. Data homogeneity across cohorts and validation through 250+ published studies ensure high quality of the datasets.

We selected parishes containing both pre- and post-vaccination cohorts (see Figure A2, Appendix A). Generation 1 includes those born 1790–1820; Generation 2 (their children) were born 1805–1865 (mothers aged 15–45); Generation 3 (grandchildren) were born 1820–1910. We tracked 141,067 individuals across three full life cycles until death, out-migration, or age 100.

Figure A3 (Appendix A) shows below-10 mortality rates by cause (smallpox, other infectious, non-infectious) from 1790–1920. Causes of death are coded using ICD-10, based on expert encoding. While smallpox declined post-vaccination and child mortality decreased slightly, cohort life expectancy at age two surged 14 years for those born 1801–1820 versus 1790–1800, continuing to grow in subsequent generations.

***Individual Vaccination.*** Our key treatment variable is whether Generation 1 individuals received smallpox vaccination by age 2. Microdata show vaccination typically occurred in the first years of life (median: 2.04 years for post-1801 cohorts). We avoided continuous vaccination dates due to potential imprecision, as changing frequencies suggest recording inaccuracy.

Controls include never-vaccinated individuals (those with natural immunity from surviving smallpox or lacking both vaccination and immunity). Vaccination status derives from priests' church book marks recorded during censuses and life events. Later vaccination ages may reflect family absence rather than initial vaccination dates, reinforcing our exclusion of migrant families and children vaccinated after age 2 from controls. Natural immunity cases are too rare for separate analysis.

We verified measurement accuracy by comparing microdata vaccinated counts with aggregated parish censuses. Medical reports confirm these priest-reported census records as most accurate (Riksarkivet 1796-1820). Despite some lost parish books potentially affecting data, counts matched perfectly.

***Lifetime Outcomes.*** Data include death or out-migration dates. Southern Sweden records link to the Swedish Death Index (Släktforskarförbund 2019); central/northern Sweden has 54% death date coverage. Our longevity indicator counts years lived post-age-two, also indicating latest observation time. Results remain consistent when treating these as censored observations.

Occupational data from church books, poll-tax, and examination registers serve as outcomes, controls, and IV construction. Annual records are coded into historical social stratification as continuous occupational scores (Lambert et al. 2013), enabling cross-cohort comparisons. For the instrument, we selected specific occupations (church assistants, musicians) and augmented data with parish/county clergy counts.

Additional outcomes include disability onset (blindness, deafness, mental/behavioral disorders, general weakness) from church records for northern/central Sweden, and literacy data until 1870 for northern parishes (Generation 1). Annual literacy tests involved confidential priest talks, often followed by social gatherings.

Offspring outcomes include life expectancy, disability-free life expectancy, and occupational scores. We exclude literacy effects due to near-perfect positive collinearity with parental vaccination status.

### *2.2 Selective Attrition*

Our historical longitudinal dataset is prone to selective migration but not false linkage. Below we highlight data quality and our mitigation strategies.

Digitized family links and demographic events ensure high quality. Swedish household-based parish registers continuously updated demographic events, maintaining family units on single pages while cohabiting. Records were updated from event registers as new events occurred. This double-bookkeeping system enables accurate individual tracking (e.g., during out- and in-migration). As Sweden's official registration system until 1990, these records provide complete population coverage regardless of church affiliation. Table A1 (Appendix A) confirms digitization quality by showing no significant associations between vaccination status and birth seasons—a common inaccuracy marker.

Swedish sources only record individuals within parish borders. For our population (born 1790–1910), two strategies mitigate this: 10% linked to national death records, and data collected from neighboring parish clusters with within-cluster tracking. This improved coverage since most movements occurred within 15 kilometers, with people relocating to the same or nearby parishes rather than long distances (Dribe 2003). Median leaver age is 18.1 years, coinciding with youth household departure for work, likely in nearby non-sampled parishes.

Generations 2 and 3 include all children of individuals observed in the study area, including return migrants. We observe offspring from 37% of Generation 1 individuals. Figure A4 (Appendix A) shows observable characteristic differences between Generation 1 individuals with and without observed children. Likelihood of having observed children is uncorrelated with socio-economic traits (family occupational scores, literacy, maternal marital status), suggesting we aren't selectively observing healthier children. Birth year association is marginally significant. Substantial parish disparities exist but sources are unclear, unlinked to geography (south-north, urban-rural). Importantly, Generations 2 and 3 span all 70 parishes and 31 cohorts studied in Generation 1.

Selective migration bias direction is difficult to predict. Resource-limited families (landless, large families, widows) lacked means for long-distance moves; farmers stayed due to extensive local networks and resource access (Vikström, Marklund, and Sandström 2016). Perhaps the only coherent out-migration driver was high grain prices, negatively impacting populations by making food unaffordable (Dribe 2003). This factor's appeal is its exogeneity to pre-migration individual decisions, serving as a migration predictor and addressing selective migration bias.

We correct for selective migration using the Heckman procedure (Heckman 1979). First-stage probit models probability of estimation sample inclusion as a function of rye prices (most common grain) and birth month (common accuracy predictor), interacting these with birth year dummies following a DID structure (Sant'Anna and Zhao 2020). Rye prices from Jörberg (1972) are set based on median cohort ages at 18: 1825 (Generation 1), 1860 (Generation 2), 1910 (Generation 3). We predict inverse Mills ratios for each individual, including them as covariates in estimation models. Alternative inverse probability weighting (weights from same factors; (Weuve et al. 2012)) yielded nearly identical results. We report Heckman-adjusted estimates for each outcome and generation.

## 3. Empirical Strategy

### *3.1 Selection into Vaccination*

We begin with the analysis of selection for vaccination against smallpox by the age of two for Generation 1. From the microdata, we obtained various background characteristics of the individual, measuring parental wealth (occupational score and marital status), literacy, parenting style (survival history of the family and death of a sibling due to an external cause), as well as the year and parish of birth. Figure A5 in Appendix A presents the OLS estimates for these variables. The results show that most variables measuring family conditions correlate weakly or not at all with the probability of children being vaccinated. For instance, paternal literacy and church attendance does not influence the probability, and a one-standard deviation change in paternal occupation score increases the probability by only 0.012 percentage points. These results align well with the fact that vaccination was free for parents and did not face opposition in Sweden.

The results also indicate that differences in a child's vaccination status primarily stem from the parental parish of residence in the first years of the child's life. The differences across parishes in the proportion of children vaccinated by the age of two vary between -1.1 to 0.16 percentage points compared to the baseline. Previous research has shown that the availability of vaccinators, such as the ratio of clergy, church assistants, and church musicians per population, explains such geographical differences (Sköld 1996). Another significant factor is the year of birth, as vaccination was first introduced in 1801, and the vaccination rate steadily rose in the subsequent years. The findings therefore suggest that the factors driving vaccination of a child by the age of two appear at the regional and cohort level.

Even with no indication of selection into vaccination at the family level based on observables, selection could still appear from unobservable factors. For instance, parishes with a higher share of vaccinated children may also be characterized by higher levels of trust to authorities, which, in turn, are eager to implement health policies that benefit parishes' residents, such as employ licensed midwives or practice isolation of sick residents before it became a widespread health measure (Lazuka, Quaranta, and Bengtsson 2016). On the individual level, families that decide to vaccinate their own child may also be more cooperative, for instance, and such social norms could affect the child's future outcomes and the outcomes of the next generations regardless of initial vaccination. To address selection into treatment for Generation 1, we apply three methods: mother fixed effects, DID, and SSIV. As we show below, we need all three methods to ensure that we obtain quasi-random variation in individual vaccination status that we can use to explore the intergenerational transmission of this vaccination's benefits to the outcomes of Generations 2 and 3.

### *3.2 Mother Fixed Effects*

As a first strategy to address unobservable selection into a child's smallpox vaccination, we employ mother fixed effects. Mother fixed effects will allow us to compare biological siblings, one of whom received the treatment while the other did not. This effectively removes all fixed, unobservable family-related factors that may affect both the treatment and long-term outcomes in life. The model is as follows:

$$(1)\ Y_{iprt} = \beta Vaccinated_{iprt} + \mu_m + (\eta_t + \gamma_p + \delta_{rt} + \boldsymbol{X}_{i(p)t}\Gamma) + \varepsilon_{irpt}$$

The index $i$ denotes individuals in Generation 1, $\mu_m$ is a set of mother fixed effects. The dependent variable, $Y_{iprt}$, is an individual's outcome, such as years lived after the age of 2, disability-free years lived after the age of 2, literacy, and occupational score. In additional specification, we will add a set of controls to effectively control for time-varying factors at the family and area levels, such as changes in parish of residence, local health policies, or family-specific disease events. Vaccination here can be treated as a positive external shock to one sibling but not the other, occurred because at that time family resided in the parish where vaccination was provided or the child was born when vaccine became available, even though the family decisions were orthogonal to the decision to vaccinate both of their children.

### *3.3 DID Approach*

Our second strategy to deal with selection is to adapt a DID design previously used to identify the outcomes of exposure to certain infectious diseases targeted by nationwide rapid interventions (Lazuka 2020, Ager, Hansen, and Jensen 2017). We exploit two sources of variation: (i) the varying exposure of different cohorts to the introduction of the smallpox vaccine in 1801, and (ii) the relatively larger benefits for individuals born in parishes with greater availability of church assistants and musicians compared to those born in parishes with lower availability. As explained in Section 3.2, our sample includes individuals who were either vaccinated by age 2 or never vaccinated; thus, the arrival of the vaccine in 1801 serves as a clear watershed, nearly perfectly distinguishing treated and untreated cohorts. Pre-vaccine availability of church assistants and musicians serves as a measure of a parish's readiness to perform vaccinations (or local demand)—the higher the availability, the greater the vaccine uptake. To capture pre-treatment variation, we thus do not use the smallpox infection rate, as our focus is ultimately on the impact of the vaccination itself, not the disease.

We estimate the following model:

$$(2)\ Y_{iprt} = \beta PostXC_{post\text{-}1,p} + \eta_t + \gamma_p + \delta_{rt}\ (+ \boldsymbol{X}_{i(p)t}\Gamma) + \varepsilon_{irpt}$$

The index $i$ denotes individuals in Generation 1, $p$ denotes parishes, and $t$ is cohort. Post refers to individuals born from 1801 onward and belonging to Generation 1. For Generation 1, we analyze a panel of 31 cohorts, born between 1790 and 1820, across 70 parishes. $C_{post\text{-}1,p}$ is pre-1801 availability of church assistants and musicians at the parish level. $\eta_t$ is cohort (i.e., year of birth) fixed effects. $\gamma_p$ is parish fixed effects. $r$ denotes six geographic regions (counties), and we include region-by-cohort fixed effects, $\delta_{rt}$, into the main model to capture the potential impact of the divergent development of regions on the outcomes: In the period of study the county was governed by a county governor with sole responsibility, including the control of vagrants and hospitals, and the stipulation of smallpox vaccination campaign (Sköld 2000). The dependent variable, $Y_{iprt}$, is an individual's outcome, such as years lived after the age of 2, disability-free years lived after the age of 2, literacy, and occupational score.

Pre-1801 vaccinator availability is measured as the average number of church assistants and musicians in each parish from 1790 to 1800, as shown in Figure 2. To facilitate interpretation of the estimates, we normalized these availability indicators by dividing by the interquartile range (9.6 assistants/ musicians), representing the difference between the 75th and 25th percentiles of the indicator distribution. The resulting indicator is a continuous variable ranging from 0 to 2, with values at both endpoints covering more than 15% of the distribution, while the remaining values fall in between. Note that the availability indicator starts at 0 because it does not account for priests: each parish was part of a *pastorat* and had a priest serving.

[Figure 2 is about here]

The parameter $\beta$ in Equation 1 captures the (reduced-form, intention-to-treat) effect of smallpox vaccination. We expect $\beta$ to be positive, as a higher availability of church assistants and musicians should lead to increased vaccination rates and, ultimately, better outcomes. Following recent methodological developments in the DID literature (review of Roth et al. 2023), we impose the following assumptions for $\beta$ to have a causal interpretation: no anticipation, the stable unit treatment value assumption, and the conditional parallel trends assumption. The no anticipation assumption is satisfied due to the sudden arrival of the vaccine and the community's lack of prior knowledge of its benefits (Sköld 2000). The stable treatment value assumption is ensured by the parishes' regulated vaccine distribution process: the vaccine was provided to priests in quantities sufficient to vaccinate all parishioners and then distributed locally by vaccinators (Sköld 1996).

Since the parallel trends assumption could be potentially violated in our setting, we impose it conditionally, controlling for group-specific pre-trends. We assume that only groups of parishes with similar pre-treatment characteristics must exhibit parallel pre-trends. First, Equation 1 already includes county-by-cohort fixed effects. In an additional specification, we also introduce interactions between cohort dummies and group-level observable characteristics (measured at the parish or individual level), in Equation 1 denoted as $\boldsymbol{X}_{i(p)t}$.

Among group-level characteristics, we include factors that influence families' decisions to vaccinate, related to the changing behavior and outcomes of parish residents (as discussed in Section 4.1). We also consider local measures (i.e., at the group of parishes level) of wealth, religiosity, and health policies to capture regional differences in responses to the mandatory vaccination campaign. Specifically, we include the number of midwives (interacted with cohort dummies) to account for healthcare development and the composition of vaccinators in the area; the smallpox death rate to control for demographic and disease conditions; the share of urban population and university students per capita to account for urbanization and social progress; the number of priests to control for religiosity; and the price of rye as an economic development indicator. To address potential correlations in local shocks, we cluster standard errors at the parish level in each specification.

Finally, to detect non-parallel pre-trends and examine the effect dynamics, we will estimate an event-study model:

$$(3)\ Y_{iprt} = \beta_t^{pre} \sum_{t=1791}^{1799} C_{post\text{-}1,p} + \beta_t^{post} \sum_{t=1801}^{1820} C_{post\text{-}1,p} + \eta_t + \gamma_p + \delta_{rt} + \varepsilon_{irpt}$$

In this model, all terms are defined as before. The coefficients $\beta_t^{pre}$ a are estimated for pre-vaccine cohorts by interacting the availability indicator $C_{post\text{-}1,p}$ (i.e., the same as in Equation 1) (as in Equation 1) with each cohort born before 1801, using the 1790 and 1800 cohorts as reference categories. The coefficients $\beta_t^{post}$ are estimated for each post-vaccine cohort.

### *3.4 SSIV Approach*

#### *3.4.1 Intuition and Equations*

The smallpox vaccination evolved gradually and even stepwise from 1801 across Swedish parishes; therefore, we adopt an SSIV approach that allows us to capture exogenous variation in children's smallpox vaccination in Generation 1, better fitting its nonlinear progression. Our ultimate goal is to use this variation to explore its impact on the outcomes of Generations 2 and 3.

Following a SSIV methodological literature, we use the instrument's formula that best describes the impact of the shock (Borusyak, Hull, and Jaravel 2022). Our instrument is $C_{p(t\text{-}1)}$ x $C_t$: it is based on the interaction between the number of church assistants and church musicians at the parish level and their ratio at the country level. Previously, in a DID approach, we defined $C_{post\text{-}1,p}$ as the pre-1801 availability of church assistants and musicians at the parish level. With the SSIV approach, we aim to measure availability for each cohort, introducing $C_{p(t\text{-}1)}$ as the number of church assistants for each parish in the previous year (cohort). $C_t$ represents the national ratio of church assistants and musicians compared to the previous year, capturing the yearly progression of vaccination as mandated by national laws (in contrast to the DID approach, where $C_t$ was essentially a binary indicator turning to 1 for post-vaccine cohorts). The interaction term mirrors the logic of the SSIV method applied to panel data, where $C_{p(t-1)}$ serves as shares and $C_t$ as national shifts (shocks).

Figure 3 presents the development of the national number of church assistants and musicians between 1790 and 1820, along with the rate of change. As shown, the rate of change captures large shifts in vaccine supply in the years 1805 and 1815, coinciding with national laws mandating the employment of church assistants for vaccination and the mandatory vaccination of young children. There is a small shift in the pre-vaccine years as well, which, however, would not impact the capture of children's smallpox vaccination since no children were vaccinated in those years.

[Figure 3 is about here]

We plot the interacted instrument for each year and parish (covering the years 1790-1820 and 70 parishes) in Figure C1 Appendix C. To facilitate interpretation, we have rescaled the instrument using its interquartile range between the 5th and 95th percentiles (14 persons or units). As a result, the rescaled instrument is a continuous variable ranging from 0 to 4.2. For nearly every parish, we observe that the size of the instrument increases after 1801, reflecting local demand and the availability of the vaccine. For half of the sample, the instrument's value is 0; we retain these observations as the null group, which reflects the never-treated group. To remind, the parishes with zero values of the instrument did not employ church assistants and musicians, but they had priests and other vaccinators to perform vaccination. Due to the identifying assumptions that we describe below, we would like to capture only the variation due to church assistants and musicians as vaccinators. The overall ranking of counties based on this instrument's quantity perfectly aligns with the county ranking observed across the entire country (Sköld 1996).

For Generation 1, our first- and second-stage equations are as follows:

$$(4)\ Vaccinated_{iprt} = \alpha(C_{p(t-1)} \text{ x } C_t) + \boldsymbol{X}_{i(p)t}\Gamma + \eta_t + \gamma_p + \delta_{rt} + \varepsilon_{irpt}$$

$$(5)\ Y_{iprt} = \beta \widehat{Vaccinated}_{iprt} + \boldsymbol{X}_{i(p)t}\Gamma + \eta_t + \gamma_p + \delta_{rt} + v_{irpt}$$

The $C_{p(t-1)}$ x $C_t$ term is the interacted instrument, representing the number of church assistants and musicians in each parish in the previous year multiplied by the national rate of change in the number of church assistants and musicians. $Vaccinated_{iprt}$ indicates the individual smallpox vaccination status before age 2 for Generation 1. All other terms are defined as previously.

We will also present the results of the reduced form because they have a clear interpretation as an intention-to-treat effects. The reduced-form equation is as follows:

$$(6)\ Y_{iprt} = \acute{\alpha}(C_{p(t-1)} \times C_{rt}) + \boldsymbol{X}_{i(p)t}\Gamma + \eta_t + \gamma_p + \delta_{rt} + u_{irpt}$$

For Generation 2 and 3, the second-stage equation is as follows:

$$(7)\ Y_{jiprt} = \beta \widehat{Vaccinated}_{jiprt} + \boldsymbol{X}_{ji(p)t}\Gamma + \eta_t + \gamma_p + \delta_{rt} + v_{jirpt}$$

The index $j$ denotes children (Generation 2) and grandchildren (Generation 3). We stack individual observations for the sample of mothers and fathers (grandmothers and grandfathers from the mother's and father's side) because, as we find, the effects are similar regardless of the parent's (grandparent's) gender.

For Generation 1, we analyze a panel of 31 cohorts, born between 1790 and 1820, for 70 parishes. Generation 2 consists of biological children of Generation 1 (and who are born between 1805 and 1865), and Generation 3 consists of biological grandchildren of Generation 1 (and who are born between 1820 and 1910). Equation 3 models the outcomes of children and grandchildren, $Y_{jiprt}$, as a function of variables from Generation 1, aiming to estimate the total vaccine effect transmitted across generations. In Section 6.2, we will further explore the mechanisms behind this transmission by applying causal mediation analysis, incorporating the characteristics of children and grandchildren.

### *3.4.2 Identifying Assumptions*

The estimates of the instrument's effects on the vaccination and the outcomes, $\alpha$ and $\acute{\alpha}$, should be positive in every equation. Their interpretation resembles that of the DID strategy, where we compare pre- and post-vaccine arrival cohorts and parishes with varying availability of church assistants and musicians, which reflects the parish's readiness to vaccinate or the local demand for vaccination. With the SSIV approach, we compare cohorts experiencing different national shifts (reflecting vaccine supply) across parishes with varying readiness or local demand, measured as changing across cohorts. For causality, $\alpha$ and $\acute{\alpha}$ must satisfy assumptions that we mentioned for DID.

However, our main interest is in obtaining causal estimates of the effects of Generation 1's individual smallpox vaccination, $\beta$. The causality requires four assumptions (Imbens 2014). If these assumptions hold, our SSIV estimates reflect the average effect for the observations that comply with the instrument, i.e., a local average treatment effect (Angrist, Imbens, and Rubin 1996). Even though smallpox vaccination is individual, our instrument captures its group variation by construction. In our setting, compliers are therefore parishes with higher proportions of vaccinated children that responded to the interacted instrument. Those parishes that did not respond to the instrument do not contribute to the estimate. We will address the assumption of instrument relevance in Section 5.3.1, and here, we will focus on the remaining three assumptions.

First, the instrument has no direct effect on the outcomes other than through its influence on the assignment to vaccination (exclusion restriction). To address this assumption, we chose to focus on church assistants and church musicians as the only subgroup of vaccinators. Historical sources highlight that Church workers were trustworthy and literate yet lacking knowledge on medicine (Sköld 1996). The law of 1804 stipulated that each parish must employ a church assistant or musician for vaccination, thereby blocking the potential monopolization of the process by doctors. Vaccinations were easy to learn, following the instructions distributed by the state and short training by the priest. As an illustration, a church musician who assisted at choirs became the first vaccinator in Kävlinge, one of the parishes in southern Sweden, and vaccinated against smallpox as his second part-time job; he did not participate in other health-related matters (Landsarkivet i Lund 1805-1827). In the neighboring parish, initially, a licensed midwife vaccinated children (Landsarkivet i Lund 1785-1857). Although the means of preventing disease were very limited in the beginning of the 19th century, some were practiced by doctors, such as cause-of-death counting and isolation of the sick, or by midwives, such as proper assistance at labour. In these two parishes, our instrument will capture the vaccination efforts of a church musician but not of a midwife.

Second, conditional on controls, we assume that there are no omitted common factors affecting both the instrument and the outcome (random assignment). In our case, the series of lagged church assistants and musicians in the parish along with the shocks in current church assistants and musicians in the region (which constitute two components of the interacted IV) are likely to be correlated with both fixed and varying characteristics of the parish (region), such as wealth or religiosity, for instance, which influence the outcomes too. In practice, this is not a serious problem for our estimates for several reasons. Parish fixed effects in the baseline specification control for all permanent factors at the parish level affecting the employment of church assistants and musicians. We also introduce families' characteristics affecting families' decision to vaccinate children interacted with cohort dummies (as in section 4.1), which account for the parish shocks related to the changing parish residents' behavior and outcomes. To identify any unobserved time-varying parish shocks, we examine pre-trends and find none, as we further elaborate on in Section 5.3.2.

In relation to the regional shocks, they similarly can reflect regional health policies, other than vaccination. The region-year of birth fixed effects in our baseline specification account for any such effects, observed and unobserved. Our analysis also introduces interactions between cohort (i.e., year of birth) dummies and local (i.e., for the group of parishes) measures of wealth, religiosity, and health policies, which capture differential responses of regions to the mandatory vaccination campaign. In particular, we include the number of midwives (interacted with cohort dummies) that will control for development of healthcare and composition of the vaccinators' group in the area; smallpox death rate will control for demographic and disease conditions; the share of urban population and university students per capita will control for the urbanization and progressivity; the number of priests will difference out the effects of religiosity; and the price of rye will control for economic development. Finally, to account for the mutual correlation of the local shocks, we cluster standard errors at the parish level.

Third, the instrument must ensure that treatment becomes a more attractive option (monotonicity). We have searched the local vaccination reports (Riksarkivet 1796-1820) to identify the possibility of "defiers", i.e., cases in which parishes *reduce* the vaccination rate when there is a positive regional employment influx of church assistants and church musicians, and increase it when the influx is negative; we have not found any such cases. Under the presence of heterogeneous treatment effects, as discussed in Borusyak, Hull, and Jaravel (2022), a causally interpretable IV estimand is guaranteed as long as the treatment (i.e., individual vaccination) is correctly specified, shares are non-negative, and the true effects of shocks on each treatment are monotone (i.e., there are no "defiers"). We have already discussed that the error in the treatment variable is unlikely in Section 3.2. The shares cannot be negative by construction.

## 4. Results for Generation 1

### *4.1 Mother Fixed-Effects Estimates of the Impact of Smallpox Vaccination*

We begin by presenting the results of the mother fixed-effects estimations, shown in Table 1. The mother fixed-effects results indicate positive and highly statistically significant effects of smallpox vaccination in childhood on both health and economic lifetime outcomes of Generation 1, regardless of specification. In the models with only mother fixed effects, vaccination by age two increases the number of remaining years lived after age two by 14 years, disability-free years by 13 years, raises the probability of possessing good literacy skills by 12 percentage points, and improves occupational scores by 4.3 units. In the models with additional controls that account for time-varying effects, the vaccination estimates adjust slightly, depending on the outcome, but remain sizable and highly statistically significant (at the 99% significance level). Regarding the Heckman-correction procedure, we find it reduces the effect on years lived to 11 years, disability-free years to 10 years, with only small adjustments for good literacy and occupational scores. Therefore, we will rely on Heckman-adjusted effect estimates in our future comparisons with DID and SSIV estimates.

[Table 1 about here]

Mother fixed-effects produce the local average treatment effects of vaccination for families that have a varying treatment status of their children (D. L. Miller, Shenhav, and Grosz 2023). To analyze a mothers' complier population, in Figure B1 Appendix B, we assess the differences between the families with and without varying vaccination status of their children. We also provide the "naïve" OLS regression results—without mother fixed effects—in Table B1 Appendix B. Mothers who choose to vaccinate some children but not all have a higher proportion of children who die and husbands who are illiterate. However, our mother fixed-effects estimates are not significantly different from those obtained through a naïve control-for-observables strategy, suggesting that heterogeneity due to socio-economic factors does not drive our mother fixed-effects results. Instead, a more significant difference emerges based on the child's year of birth. This suggests that, for mothers with varying vaccination status of their children, smallpox vaccination varies primarily due to the unavailability of the vaccine for children born before 1801, which is the variation we need.

### *4.2 DID Estimates of the Impact of Vaccine Arrival*

We then turn to the DID results for the impact of smallpox vaccine arrival, which we introduce to fully capture the intuition behind the SSIV estimates. We present the DID estimates for the Generation 1's outcomes in Table 2. Both baseline and Heckman-corrected estimates are shown for specifications following Equation 2, with and without additional controls. We also present the event-study estimates in Figure 4.

[Table 2 and Figure 4 are about here]

The DID results indicate that the arrival of the vaccine significantly improved Generation 1's lifetime outcomes. The most conservative estimate of the impact on remaining years lived at age 2 is 4.5 years, and for disability-free years lived, it is 4.4 years—both statistically significant at the 1% level. For socio-economic outcomes, the estimate is 7.2 percentage points for good literacy and 1.6 units for occupational score, each also statistically significant at the 1% level. For most outcomes, the introduction of controls for pre-trends in observable characteristics yields estimates like those in the baseline specification. Finally, the Heckman correction procedure slightly reduces the estimates for health outcomes, likely accounting for the selection of families before age 15. However, the estimates with and without Heckman correction are not statistically different from each other.

The event-study results—which demonstrate the evolution of the vaccine arrival effect across cohorts—indicate the appearance of the vaccine effect for the outcomes of post-vaccine cohorts (i.e., born from 1801 onward), coinciding with the vaccine's inception. For these cohorts, the effects remain consistently positive, with a tendency to rise. There is a slowdown around the year 1805: in this year, parishes, following the national law, began actively employing vaccinators among church assistants and musicians, which may have interacted with an effect stemming from the pre-treatment number of vaccinators. For the pre-vaccine cohorts, we do not find any significant effects, suggesting that parishes with different numbers of church assistants and musicians were on the same pre-trends and supporting our main identification assumptions. These results align with the finding that DID estimates remain stable after adding controls.

To align the DID results with the mother fixed-effects results, we calculate the indirect least squares estimates for individual smallpox vaccination. We do this by dividing the DID estimates by the proportion vaccinated among Generation 1, which is 35%. The final effects, based on the conservative estimates, are as follows: 12.8 years for remaining years lived at age 2 (4.5/0.35), 12.6 years for disability-free years lived at age 2 (4.4/0.35), 21 percentage points for good literacy (7.2/0.35), and 4.6 units for occupational score (1.6/0.35). These effect sizes closely align with our mother fixed-effects estimates, suggesting that we are capturing the effects of the same phenomenon. However, the indirect least squares provide only effect sizes, not the variation in smallpox vaccination within Generation 1 that could be used to explore the outcomes for Generations 2 and 3. For this, we implement an SSIV strategy.

### *4.3 SSIV Estimates of the Impact of Smallpox Vaccination*

#### *4.3.1 Descriptive Analysis*

We begin the SSIV analysis with the relationships between the interacted instrument and the probability of being vaccinated by the age of two (first-stage) and the lifetime outcomes (reduced-form), shown in Figure 5, Panel A. We aggregate the estimation sample by the birth cohort and plot the observations and the fitted line, weighted by the number of individuals in each cohort. To interpret, recall that we have rescaled the instrument using its interquartile range between the 5th and 95th percentiles. Therefore, in our context one unit change in the instrument means its maximum impact. The results show a strong positive association: one unit change in the interacted instrument increases the share of vaccinated children by 31%.

[Figure 5 about here]

The reduced-form relationships are shown on Figure 5 Panel B, for each lifetime outcome separately. All outcomes are positively associated with the instrument. One unit change in the instrument is associated with 1.45 more years lived and 1.25 more disability-free years lived after the age of two. The relationship with economic outcomes in adult ages is also strong: the share of individuals with good literacy skills increases by 7 percentage points and occupation score by 1.13 units, because of one unit change in the instrument. The figures are rough approximations of the first-stage and reduced-form estimates, because they do not consider control variables, important for the identification of the causal effects.

#### *4.3.2 Reduced-form and 2SLS estimates*

We turn to the SSIV estimates of the impact of vaccination by the age of two on lifetime outcomes in later ages. For completeness, we will report both baseline and Heckman-corrected estimates.

To start with, for all outcomes and specifications, the Kleibergen-Paap F-statistic, which is robust to heteroskedasticity and clustering in errors, in the first stage is close to 50, meaning that the interacted instrument yields a strong impact on the probability of being vaccinated in the first ages of life (Keane and Neal 2023). The first-stage results imply that a unit change in the interacted instrument increases the probability of being vaccinated by the age of two by 15.4 percentage points, when considering a sample of individuals for years lived as an outcome, for instance. In comparison to the results of the descriptive analysis, we observe a reduction in the impact of the instrument on the vaccination rate, due to the inclusion of the control variables necessary for causal identification.

The baseline 2SLS estimates are shown in Panel A and Heckman-corrected 2SLS estimates are in Panel B, Table 3. Regarding the impact on the outcomes, the 2SLS estimates show a positive and statistically significant impact of vaccination on health variables and an occupational score. Here the reduced-form estimates are also strong. The Anderson-Rubin 2SLS test statistic draws a similar conclusion about the impact of vaccination as the 2SLS *t*-statistic, indicating that the latter is unbiased (Keane and Neal 2023). Regarding the reduced-form effects, a one unit increase in the instrument (recall: the maximum impact) increases the number of years lived by 1.8 and the number of disability-free years by 1.7 years. The 2SLS estimate for the impact of vaccination on years lived and disability-free years lived is 12 years each. The Heckman-corrected estimates are one year larger than the baseline.

[Table 3 about here]

Turning to economic outcomes, the reduced-form impact of the instrument is 0.5 units of occupational score in adult ages, and the 2SLS impact of vaccination is 3 units, which is statistically significant at the 1% level. For literacy, the 2SLS impact of vaccination loses statistical significance but remains in a similar magnitude, with a 11-percentage-point increase in the probability of having good literacy skills in adult ages. For this outcome, our 2SLS estimates are thus not informative about the causal vaccination effect, which, however, does not imply that there is no such effect. Since we further find vaccination effects on mental disability, the absence of statistically significant SSIV results for literacy is likely due to the noisiness of the literacy measure. For the economic outcomes, the Heckman-corrected estimates are not different from the baseline.

We observe that our results are robust to the inclusion of additional parish- and time-varying controls, but the estimates tend to become less precise when more controls are added. This happens because many controls in the extended model are also strongly correlated with the vaccination variable, so adding these controls reduces variation in the vaccination variable induced by the instrument and increases standard errors. The point estimate of the vaccination effect is smaller in the instrumental-variables compared to the controlling-for-observables estimations but remains similar in statistical terms. However, even the lowest, a 10-year increase in years lived due to smallpox vaccination in the 2SLS estimations is large enough to account for most of the cohort improvements in life expectancy after the age of two.

#### *4.3.3 Parallel Pre-Trends*

It is possible to validate the shift-share instrument by examining whether observations with different exposure shares exhibit parallel trends prior to the shock (Borusyak, Hull, and Jaravel 2022). This follows from that linear instrumental-variables with shift-share as an instrument varying by year of birth and parish of birth are analogous to the DID with continuous treatment, with underlying "parallel trends" restrictions. Any evidence for significant pre-trends would signal that a set of cohorts in particular parishes with different levels of the instrument would have evolved differently from each other even in the absence of the vaccination campaign.

To assess the plausibility of the parallel trends, we create the leads of the instrument and estimate the reduced-form effects with these leads instead of the instrument. The results for the models with two sets of covariates are presented on Figure C2 Appendix C. Overall, there is no evidence of significant pre-trends in the outcomes. However, note that the evidence for no pre-trends points to the absence of the time-varying shocks, sufficient for the causality claim of the reduced-form estimates.

### *4.4 Influence of Epidemics and Other Interventions*

In this section, we explore the role that positive and negative shocks, aside from vaccination, played in generating our results.

Presumably, smallpox epidemics severely affected the cohorts in question, particularly before the introduction of the vaccine, and left behind a negatively selected group of children who serve as a comparison. If this is the case, our findings on lifetime effects should be interpreted as the combined result of the vaccine's positive impact and the benefits of avoiding the detrimental effects of epidemics. Previous research has demonstrated that airborne disease epidemics, including smallpox, left long-term scars on survivors, manifesting in lower survival rates (Quaranta 2014).

The institutional context of the vaccination campaign suggests that no beneficial interventions took place in parallel. Medicine was underdeveloped at that time, and epidemiological hospitals helpful in isolating infectious disease were built much later (Lazuka, Quaranta, and Bengtsson 2016). One potential intervention to consider was the practice of midwives—they improved maternal survival and could influence infant health, with long-term health consequences. However, previous research has found no effects on infant health prior to the acceptance of the germ theory of disease and re-education of midwives at the late nineteenth century (Lorentzon and Pettersson-Lidbom 2021; Lazuka 2018). Another intervention to contemplate is the introduction of potatoes around 1805. At that time, farmers began growing potatoes in arable fields, which increased the production of nutritious food, potentially benefiting the growth and health of the population (Lazuka, Bengtsson, and Svensson 2023).

We assess the influence of smallpox epidemics and other interventions on vaccination against smallpox and lifetime outcomes by introducing an interaction of child smallpox mortality, the ratio of midwives to population, and the quantity of potato seeds per square kilometer (observed at the parish level) with an interacted IV into Equation 3 (the reduced-form equation). Table D1 in Appendix D reports the results.

The reduced-form estimates of the interacted instrument on all outcomes remain largely unchanged after accounting for interactions with other shocks, suggesting a strong and unbiased direct effect of vaccinators' efforts on lifetime outcomes. Moreover, there is an additional beneficial effect for individuals who are positively affected by vaccination efforts and born during smallpox epidemic years, amounting to an increase of one year in lifespan and two units of occupational score. Since the early smallpox vaccine was highly effective (Hedin 1802), this additional effect likely arises from the detrimental outcomes experienced by unvaccinated individuals, whose lives were significantly impacted by the disease. Therefore, two-thirds of the effects we observe are likely due to analyzing a historical population severely affected by epidemics.

For positive cointerventions, we do not find any interactive effects. For midwives, the absence of the interaction effect aligns with the observation that midwives or doctors did not serve as the primary group of vaccinators. Regarding the impact of potatoes, the results reinforce the notion that smallpox is a disease with a low correlation to nutritional intake (Fenner et al. 1988).

## 5. Results for Generation 2 and Generation 3

### *5.1 The Effects on Longevity, Disability, and Occupational Score*

We next consider the results for Generation 2 and 3, having shown that the SSIV strategy provides us with a vaccination effect for Generation 1 that is likely causal and similar to alternative causal identification strategies. In this section, we turn to the effects of the Generation 1's childhood vaccination status on the outcomes of Generation 2 (to whom Generation 1 are parents) and Generation 3 (to whom Generation 1 are grandparents). We stack individual observations for the sample of mothers and fathers (grandmothers and grandfathers from the mother's and father's side). As explained in Section 4.4.1, we model the outcomes of Generation 2 and 3 as a function of variables from Generation 1, aiming to estimate the total vaccine effect transmitted across generations. We also provide the Heckman-corrected effects, as detailed in section 3.4.

Table 4 presents the SSIV results for the outcomes of Generation 2 and 3 in Panels A and B. Smallpox vaccination of Generation 1 improves the health lifetime outcomes of both Generation 2 and 3, and the effects are statistically significant at least at the 5% level. If to rely on the magnitude with a more extended set of controls, life expectancy at birth increases by 2.2 years for Generation 2 and 1.1 years for Generation 3. Regarding occupational score, we find the marginally statistically significant effect for Generation 2 only, with a 1-point increase on a continuous scale, while Generation 3 does not benefit from grandparents' vaccination status. The relative magnitudes of the effects on life expectancy and occupational score are around 20% and 10% for Generation 2 and 3, when compared to the effects observed in Generation 1. The increase in disability-free life expectancy—a measure of morbidity—is even more substantial, with gains of 8 years for Generation 2 and 4.3 years for Generation 3, which is more than a half of disability gain for Generation 1. As with Generation 1's results, a Heckman correction procedure updates the coefficients only slightly.

[Table 4 about here]

Narrowing the birth cohorts to individuals born close to 1845 and 1890, with these years serving as the median for Generation 2 and 3, does not alter the results. Our findings regarding the health of succeeding generations therefore suggest potential mediation (i.e., reinforcement) by other influential factors, such as health interventions against infectious disease implemented from the 1880s, for instance. The absence of transmission of socio-economic relationship to the distant generations is consistent with increased social mobility among cohorts born around the 1890s in early industrializing Sweden. We further explore these possibilities in the causal mediation analysis.

### 5.2 Mechanisms of Intergenerational Transmission

In the final section, we conduct a causal mediation analysis of the impact of Generation 1 vaccination on the outcomes of subsequent generations. While there are numerous potential mediators in these long-term relationships, our primary focus is to distinguish between factors related to nature and nurture. Specifically, we aim to determine whether smallpox vaccination induces people to transmit knowledge through nurturing and/or if there is a direct biological transmission of past environments (i.e., epigenetic inheritance) (Collado, Ortuño-Ortín, and Stuhler 2023).

We select mediators based on their relevance to nurturing and natural influences. To measure *nurture*, we analyze variables such as the childhood smallpox vaccination status of Generation 2 (and Generation 3), whether the child was assisted by a midwife at birth, and the parental occupational score. For *nature*, we use the fixed component derived from the mothers of Generation 2 and 3. Specifically, we run mother fixed effects regressions on the outcomes of Generation 2 and 3 (i.e., years lived, disability-free years lived, and occupational score) and predict a part of each outcome shared within mother for each individual—our measure of epigenetic factors.

To perform the causal mediation analysis, we follow the causal inference literature, which proposes to decompose the total effect into the natural direct effect and natural indirect effect (Imai, Keele, and Tingley 2010). In our case, the natural direct effect is comparing $Vaccinated_{iprt}$ at 1 and 0 intervening to fix the level of mediator to 0, and the natural indirect effect is comparing the effects at different levels of mediator, intervening to fix $Vaccinated_{iprt}$ at 1. Both effects are identified under the assumptions of no unmeasured confounders between the combinations of treatment, outcome, and mediator, given a set of observables. To improve the plausibility of these assumptions, we perform the analysis on a set of covariates that we used in a 2SRI model.

We follow Imai, Keele, and Yamamoto (2010) and estimate the direct and indirect natural effects by fitting a parametric regression model for the outcome of Generation 2 (Generation 3) and a parametric regression model for mediator in the following ways:

$$(9)\ Y^{2(3)}_{ijprt} = \beta Vaccinated_{ijprt} + \mu M^{2(3)}_{jprt} + \boldsymbol{X}_{i(p)t}\Gamma + \eta_t + \gamma_p + \delta_{rt} + \widehat{\varepsilon_{ijrpt}} + \nu_{ijrpt},$$

$$(10)\ M^{2(3)}_{ijprt} = \beta Vaccinated_{ijprt} + \boldsymbol{X}_{i(p)t}\Gamma + \eta_t + \gamma_p + \delta_{rt} + \widehat{\varepsilon_{ijrpt}} + \epsilon_{ijrpt},$$

where $Y^{2(3)}$ is the outcome for Generation 2 (Generation 3), and $M^{2(3)}$ is the mediator for Generation 2 (Generation 3). All other terms are defined as before. After fitting the models *(4)* and *(5)*, the method then uses simulations to calculate natural direct and natural indirect effects.

Robust standard errors are clustered at parish of birth of Generation 1 and retrieved by bootstrapping.

Table 5 displays the results for direct and mediated effects on offspring's outcomes, in relation to four mediators. To interpret the importance of the mediator, we need to multiply the estimate for the natural indirect effect by the mediator's mean. For the health outcomes, our results point to the importance of both nurturing and epigenetics. The propensity of parents to vaccinate their children against smallpox emerges as a significant, reinforcing mediator: because parents (grandparents) were vaccinated against smallpox before age two and vaccinated their children, Generation 2 and 3 experience gains of 1 and 0.8 years of life, as well as 2.8 and 2.5 disability-free years, respectively. The impact of epigenetic factors is also sizable but emerges only for Generation 2: the mean of epigenetic factors (shared life expectancy) is 0.05, meaning that the estimate for the mediated effect of epigenetic factors is 0.3 added years of life expectancy. The impact of other mediators, like parental occupational score and assistance at birth by a licensed midwife, is low or marginally statistically significant. In total, the impact of both parental vaccination and epigenetic factors fully explains the transmitted effect of the Generation 1's vaccination effect on health of subsequent generations.

[Table 5 about here]

For the occupational scores of subsequent generations, the mediated results show a different pattern. Occupational score of Generation 2 increases by 2.2 units due to parental smallpox vaccination directly. However, the natural indirect effect of parental occupational scores amounts to 0.07 per unit and is statistically significant at the 5% level. With a range of parental occupational scores of 61.8 units, the maximum mediated effect amounts to 4.5 units (61.8 x 0.072), which is even larger than the direct effect. Previously, we did not find any significant total effects of grandparental smallpox vaccination on the occupational scores of Generation 3. We now observe a substantial indirect effect through parental occupational scores: individuals belonging to Generation 3, whose grandparents were vaccinated against smallpox by age two, receive a gain of 0.09 per unit of their own occupational score, or 5.4 units at the maximum (61.8 x 0.087).

Overall, a positive shock to health of Generation 1, such as smallpox vaccination, operating through various channels, enhances both health and socio-economic outcomes for at least two more generations.

## Conclusions

In this study, we investigated whether smallpox vaccination in early childhood enabled individuals to live longer and become prosperous as adults, and whether such vaccination effects were transmitted to their two consecutive generations. To obtain the causal effects of smallpox vaccination, we applied a SSIV approach, using the fact that vaccination in Sweden was carried out by low-skilled clergy who otherwise did not perform public health duties. We leveraged unique historical microdata from different areas across Sweden, covering the full lifespans of three generations.

Our study leads us to draw several important conclusions. First, we find evidence consistent with both specific and non-specific vaccination effects. Smallpox vaccination increases the total and disability-free life expectancy of Generation 1 by 11 years and enhances their occupational achievements by 10%. Such effects emerged due to decreases in mortality from smallpox and other causes, but also appear to have reduced disability associated with ailments that could hinder human capital accumulation. Second, these effects persist across generations. Smallpox vaccination of Generation 1 increases the life expectancy of Generation 2 by 2 years and of Generation 3 by 1 year, as well as improving their occupational scores. More than half of the transmitted effects are attributed to nurture, as vaccinated individuals are more likely to vaccinate their children across generations, while epigenetic factors account for the remainder. Finally, we obtain similar results when utilizing different causal strategies, such as SSIV, DID, and mother fixed-effects. The results withstand a large number of robustness checks.

Our findings are potentially important for policy as they underscore the power of vaccination. The evidence that smallpox vaccination offers protection not only against smallpox but also against other diseases makes vaccination a powerful health intervention. Our findings, which highlight very long-term, intergenerational health and economic benefits from vaccination, suggest that the total benefits of smallpox vaccination were much larger than the existing literature suggests. We demonstrate that while a significant portion of the effects can be attributed to the analysis of a historical population severely affected by epidemics, vaccination remains beneficial in the very long term, even in milder disease environments like those seen today. Whether these findings are applicable to other vaccines is beyond the scope of this paper but remains an important topic for future research.

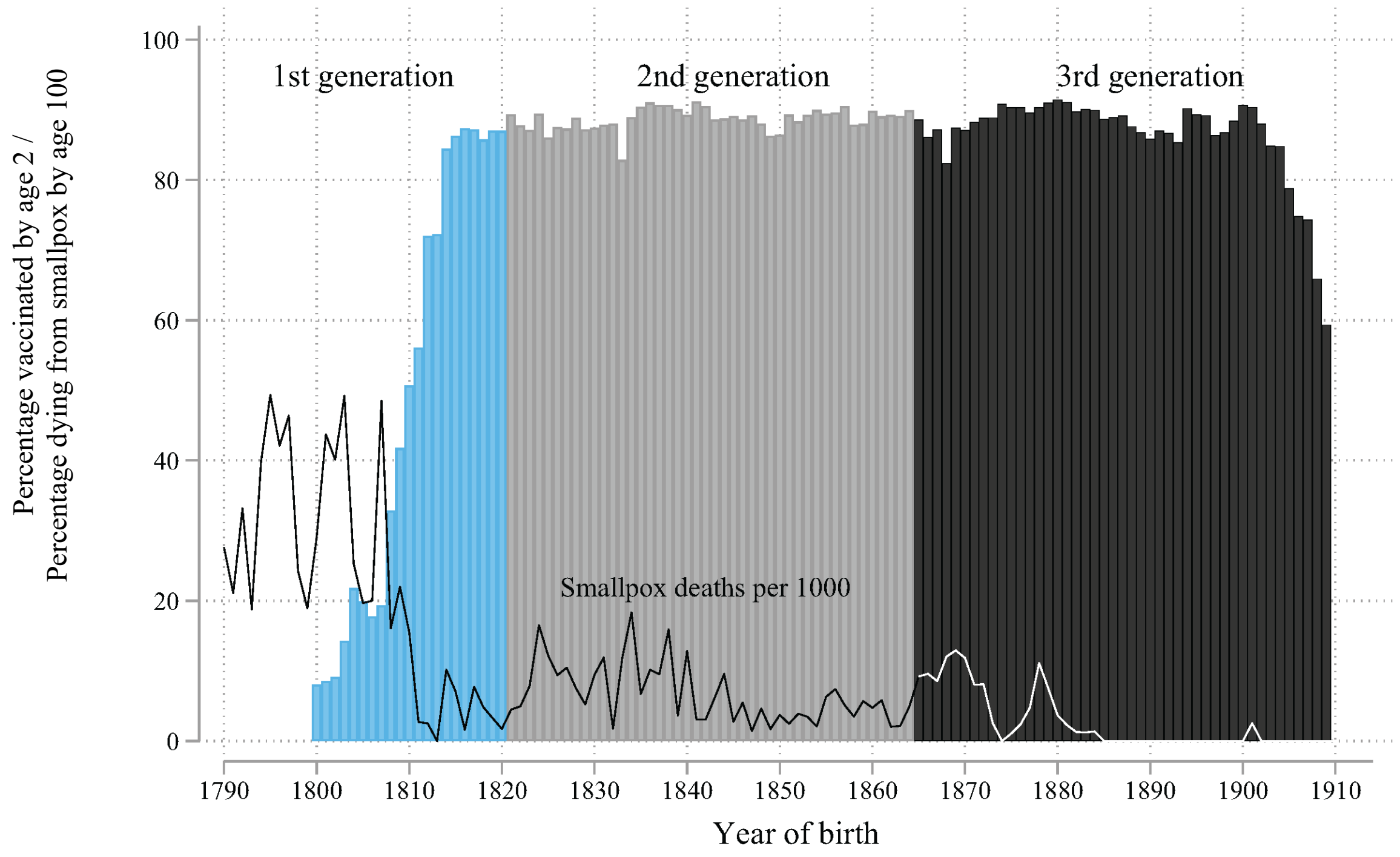

Figure 1 – Percentage of vaccinated by the age of 2 (bars) and of dying from smallpox by age 100 (line), cohorts 1790–1910.

*Source*: own calculations based on the estimation sample.

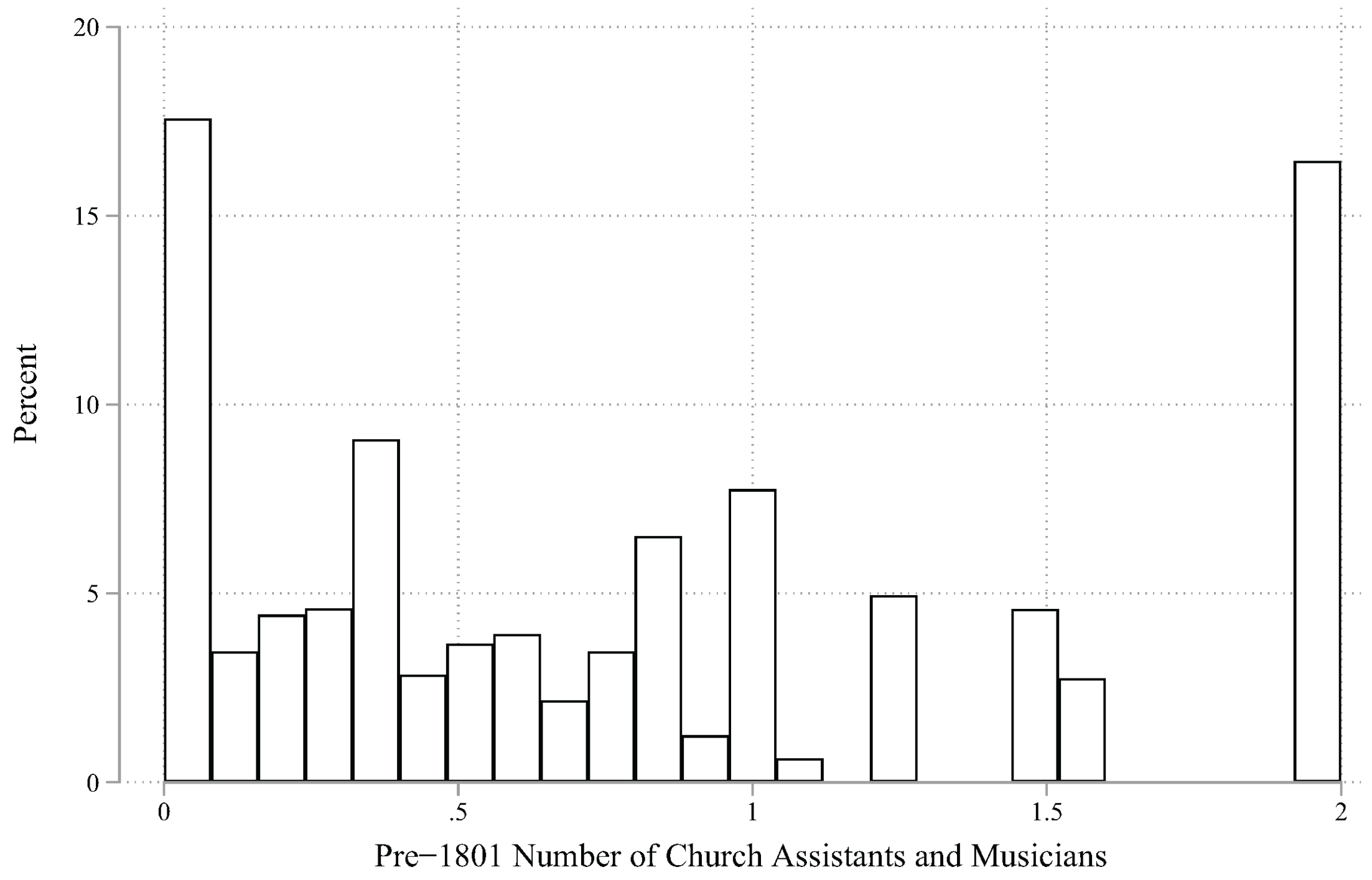

Figure 2 – Distribution of the pre-1801 church personnel availability indicator by its values, parish-level indicators.

*Source*: own calculations based on the estimation sample.

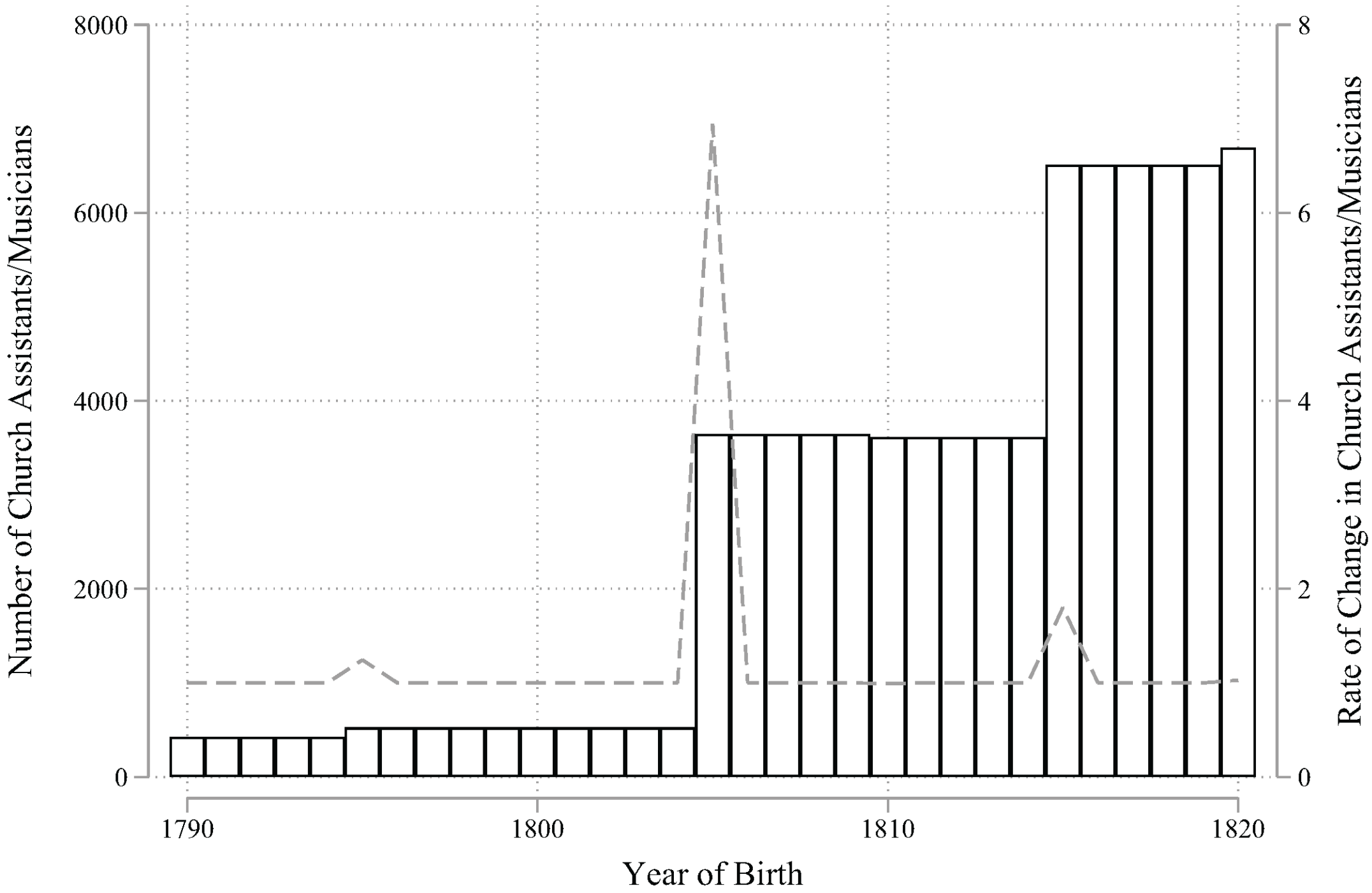

Figure 3 – The number and rate of change in church assistants and musicians in Sweden, 1790–1820.

*Source*: own calculations based on TABVERK.

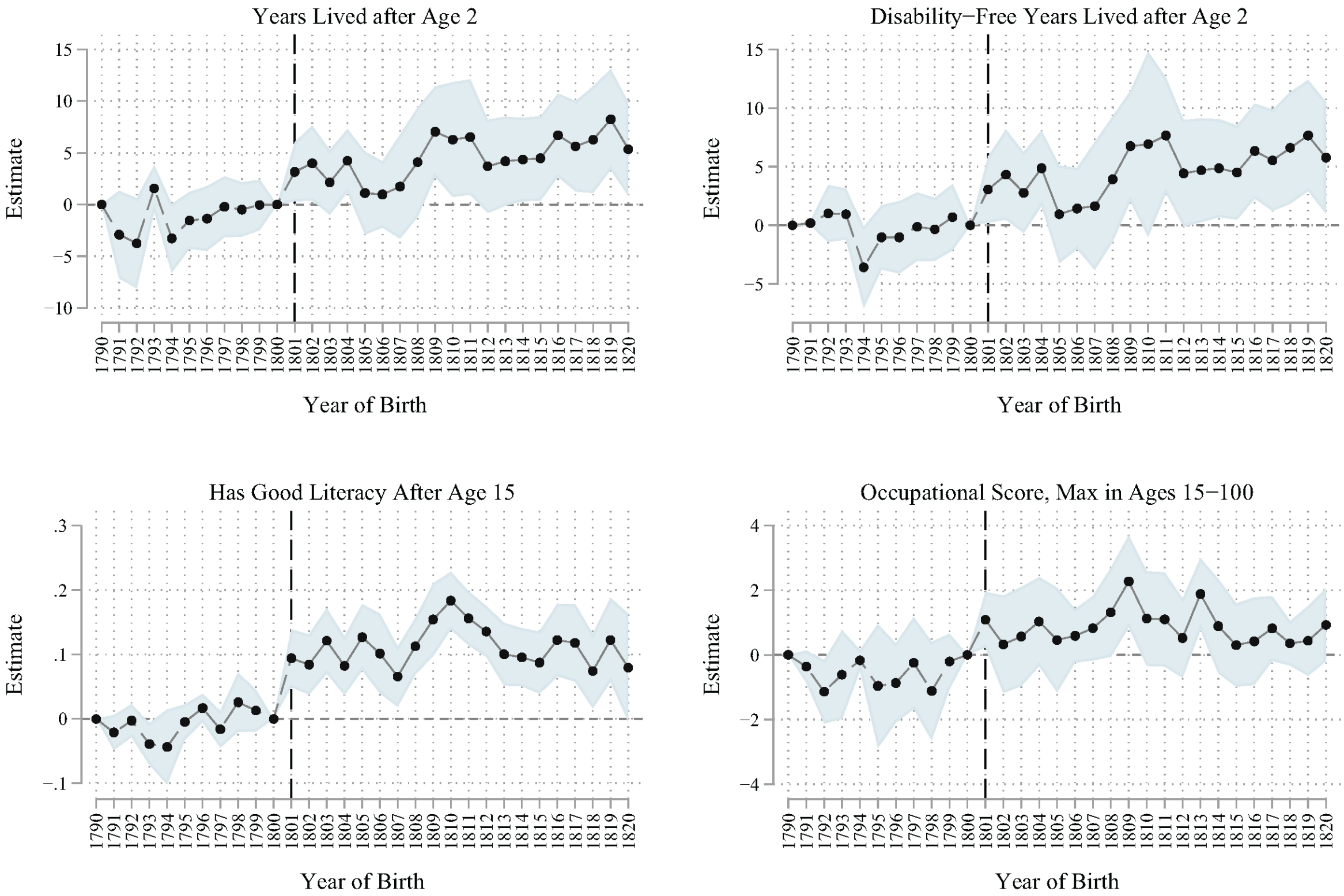

Figure 4 – Event-study and 95% confidence interval estimates of the effects of the arrival of the smallpox vaccine in 1801 on lifetime outcomes of Generation 1, using $C_{post\text{-}1}$ as a treatment intensity (pre-1801 availability of church assistants and musicians at the parish level).

*Note*: Estimated according to Equation 2.

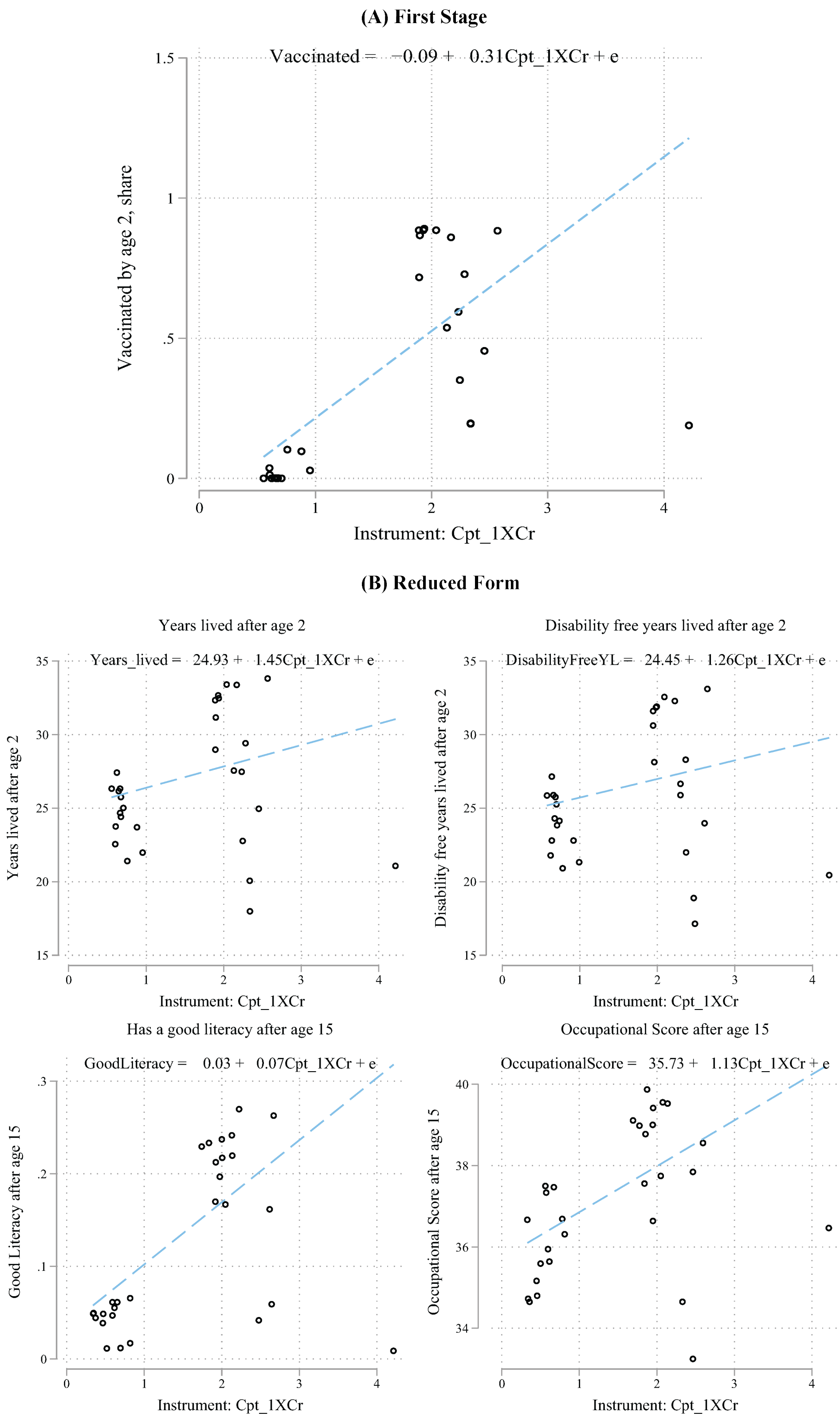

Figure 5 – First stage and reduced form: The relationship between the SSIV ($C_{p(t-1)}$ x $C_r$) and the share vaccinated by age 2 (Panel A) and the individuals' lifetime outcomes (Panel B), average by birth cohort weighted by the number of individuals.

*Source*: own calculations from the estimation sample.

Table 1 – The effect of smallpox vaccination on lifetime outcomes of Generation 1: Mother fixed effects

| | Remaining years lived at age 2 | | Disability-free years lived at age 2 | | Has good literacy, after age 15 | | Occupational score, max in ages 15-100 | |
|---|---|---|---|---|---|---|---|---|
| | (1) | (2) | (3) | (4) | (5) | (6) | (7) | (8) |
| **Panel A: Baseline estimates** | | | | | | | | |
| Vaccinated by age 2 | 13.585*** | 18.495*** | 13.265*** | 17.746*** | 0.118*** | 0.0769*** | 4.273*** | 4.474*** |
| | (0.225) | (0.349) | (0.226) | (0.352) | (0.00551) | (0.00779) | (0.109) | (0.359) |
| Rsq | 0.074 | 0.173 | 0.071 | 0.155 | 0.062 | 0.234 | 0.0025 | 0.142 |
| | | | | | | | | |
| **Panel B: Heckman-corrected estimates** | | | | | | | | |
| Vaccinated by age 2 | 11.067*** | 18.386*** | 10.286*** | 17.806*** | 0.0814*** | 0.0646*** | 4.648*** | 4.499*** |
| | (0.233) | (0.346) | (0.233) | (0.350) | (0.00536) | (0.0074) | (0.205) | (0.358) |
| Rsq | 0.091 | 0.185 | 0.096 | 0.173 | 0.148 | 0.278 | 0.0790 | 0.141 |
| Observations | 43,450 | 43,450 | 42,021 | 42,021 | 28,614 | 28,614 | 30,806 | 30,806 |
| Mother FEs | Yes | Yes | Yes | Yes | Yes | Yes | Yes | Yes |
| Parish of birth FEs | No | Yes | No | Yes | No | Yes | No | Yes |
| Year of birth FEs | No | Yes | No | Yes | No | Yes | No | Yes |
| Region of birth x Year of birth FEs | No | Yes | No | Yes | No | Yes | No | Yes |
| Families' Xs x Year of birth FEs | No | Yes | No | Yes | No | Yes | No | Yes |
| Parish of birth Xs x Year of birth FEs | No | Yes | No | Yes | No | Yes | No | Yes |

*Note:* Observations are individuals. Panel A reports the estimates based on Equation 1. Panel B reports the estimates with a Heckman correction (see section 3.4 for the procedure). The controls included are indicated in the table by Yes and No. "Families' Xs" include child characteristics at birth: sex, paternal occupational score, maternal occupational score, paternal literacy, maternal literacy, proportion of non-surviving children in the family, maternal marital status, the presence of siblings deceased due to external or unknown causes. "Parish Xs" include time-varying parish of birth characteristics: the number of midwives, the number of priests, smallpox death rate, university students per capita, price of rye, and the share of urban population.

*** p<0.001, ** p<0.01, * p<0.05

Table 2 – The effect of the arrival of the smallpox vaccine in 1801 on lifetime outcomes of Generation 1: DID estimates using $C_{post-1}$ as a treatment intensity

| | Remaining years lived at age 2 | | Disability-free years lived at age 2 | | Has good literacy, after age 15 | | Occupational score, max in ages 15-100 | |
|---|---|---|---|---|---|---|---|---|
| | (1) | (2) | (3) | (4) | (5) | (6) | (7) | (8) |
| **Panel A: Baseline estimates** | | | | | | | | |
| $PostXC_{post-1,p}$ | 4.573** | 4.937** | 4.442** | 4.647** | 0.0976** | 0.0715** | 1.590*** | 1.666*** |
| | (1.513) | (1.501) | (1.540) | (1.565) | (0.0248) | (0.0291) | (0.419) | (0.645) |
| R sq | 0.123 | 0.130 | 0.115 | 0.128 | 0.427 | 0.433 | 0.122 | 0.132 |
| | | | | | | | | |
| **Panel B: Heckman-corrected estimates** | | | | | | | | |
| $PostXC_{post-1,p}$ | 4.463** | 4.558** | 4.197** | 4.187** | 0.0967** | 0.0716** | 1.591*** | 1.668*** |
| | (1.509) | (1.494) | (1.535) | (1.565) | (0.0235) | (0.0293) | (0.427) | (0.646) |
| R sq | 0.137 | 0.143 | 0.131 | 0.143 | 0.427 | 0.433 | 0.122 | 0.132 |
| Individuals | 43,450 | 43,450 | 42,021 | 42,021 | 28,614 | 28,614 | 30,806 | 30,806 |
| Parish of birth fixed effects | Yes | Yes | Yes | Yes | Yes | Yes | Yes | Yes |
| Year of birth fixed effects | Yes | Yes | Yes | Yes | Yes | Yes | Yes | Yes |
| Region of birth x Year of birth fixed effects | Yes | Yes | Yes | Yes | Yes | Yes | Yes | Yes |
| Families' *X*s x Year of birth fixed effects | No | Yes | No | Yes | No | Yes | No | Yes |
| Parish of birth *X*s x Year of birth fixed effects | No | Yes | No | Yes | No | Yes | No | Yes |

*Note:* Panel A reports the estimates based on Equation 2. Panel B reports the estimates with a Heckman correction (see section 3.4 for the procedure). The controls included are indicated in the table by Yes and No. "Families' Xs" include child characteristics at birth: sex, paternal occupational score, maternal occupational score, paternal literacy, maternal literacy, proportion of non-surviving children in the family, maternal marital status, the presence of siblings deceased due to external or unknown causes. "Parish Xs" include cohort-varying parish of birth characteristics: the number of midwives, the number of priests, smallpox death rate, university students per capita, price of rye, and the share of urban population. Standard errors are clustered at the parish-of-birth level.

*** p<0.001, ** p<0.01, * p<0.05

Table 3 – The effect of smallpox vaccination on lifetime outcomes of Generation 1: SSIV estimates with $C_{p(t-1)}$ x $C_t$ as an instrument (2SLS)

| | Remaining years lived at age 2 | | Disability-free years lived at age 2 | | Has good literacy, after age 15 | | Occupational score, max in ages 15-100 | |
|---|---|---|---|---|---|---|---|---|
| | (1) | (2) | (3) | (4) | (5) | (6) | (7) | (8) |
| **Panel A: Baseline 2SLS Estimates** | | | | | | | | |
| I: First-Stage Estimates (on Vaccinated by age 2) | | | | | | | | |
| $C_{p(t-1)}$ x $C_t$ | 0.154*** | 0.163*** | 0.150*** | 0.155*** | 0.181*** | 0.177*** | 0.187*** | 0.196*** |
| | (0.0351) | (0.0279) | (0.0349) | (0.0267) | (0.0396) | (0.0295) | (0.0392) | (0.0311) |
| II: Reduced-Form Estimates | | | | | | | | |
| $C_{p(t-1)}$ x $C_t$ | 1.768** | 1.869** | 1.667*** | 1.489** | 0.0192 | 0.0161 | 0.564** | 0.711** |
| | (0.509) | (0.667) | (0.463) | (0.535) | (0.0278) | (0.0312) | (0.167) | (0.212) |
| R sq | 0.109 | 0.118 | 0.100 | 0.115 | 0.330 | 0.347 | 0.169 | 0.205 |
| III: 2SLS Estimates | | | | | | | | |
| Vaccinated by age 2 | 11.483*** | 11.471** | 11.791*** | 11.041** | 0.106 | 0.091 | 3.009** | 3.617** |
| | (3.172) | (3.664) | (3.248) | (3.516) | (0.0691) | (0.0518) | (1.079) | (1.129) |
| R sq | 0.154 | 0.162 | 0.145 | 0.158 | 0.330 | 0.348 | 0.141 | 0.171 |
| Kleibergen-Paap F-statistic | 51.999 | 52.999 | 48.001 | 48.999 | 45.999 | 45.999 | 52.001 | 52.999 |
| Anderson-Rubin F-statistic | 6.940 | 3.290 | 7.530 | 6.860 | 0.600 | 0.850 | 6.450 | 4.352 |
| | | | | | | | | |
| **Panel B: Heckman-corrected 2SLS estimates** | | | | | | | | |
| Vaccinated by age 2 | 12.912*** | 12.257** | 12.428*** | 12.999*** | 0.104 | 0.102 | 3.262*** | 3.049** |
| | (3.044) | (4.468) | (2.168) | (2.374) | (0.0679) | (0.0718) | (0.919) | (1.201) |
| R sq | 0.152 | 0.158 | 0.143 | 0.156 | 0.348 | 0.350 | 0.171 | 0.186 |
| Individuals | 43,450 | 43,450 | 42,021 | 42,021 | 28,614 | 28,614 | 30,806 | 30,806 |
| Parish of birth fixed effects | Yes | Yes | Yes | Yes | Yes | Yes | Yes | Yes |
| Year of birth fixed effects | Yes | Yes | Yes | Yes | Yes | Yes | Yes | Yes |
| Region of birth x Year of birth fixed effects | Yes | Yes | Yes | Yes | Yes | Yes | Yes | Yes |
| Families' *X*s x Year of birth fixed effects | No | Yes | No | Yes | No | Yes | No | Yes |
| Parish of birth *X*s x Year of birth fixed effects | No | Yes | No | Yes | No | Yes | No | Yes |

*Note:* Panel A reports the estimates based on Equations 3, 4, and 5. Panel B reports the estimates with a Heckman correction (see section 3.4 for the procedure). "Families' Xs" include child characteristics at birth: sex, paternal occupational score, maternal occupational score, paternal literacy, maternal literacy, proportion of non-surviving children in the family, maternal marital status, the presence of siblings deceased due to external or unknown causes. "Parish Xs" include time-varying parish of birth characteristics: the number of midwives, the number of priests, smallpox death rate, university students per capita, price of rye, and the share of urban population. Standard errors are clustered at the parish-of-birth level.

*** p<0.001, ** p<0.01, * p<0.05

Table 4 – The effect of smallpox vaccination of Generation 1 on lifetime outcomes of Generation 2 and 3: SSIV estimates with $C_{p(t-1)}$ x $C_t$ as an instrument for Generation 1 (2SLS)

| | Remaining years lived at birth | | Disability-free years lived at birth | | Occupational score, max in ages 20-100 | |
|---|---|---|---|---|---|---|
| | (1) | (2) | (3) | (4) | (5) | (6) |
| **(A) Generation 2** | | | | | | |
| Parent Vaccinated | 1.171** | 2.204*** | 6.836*** | 8.015*** | 1.531** | 1.099* |
| | (0.401) | (0.652) | (1.517) | (2.008) | (0.599) | (0.556) |
| R sq | 0.0564 | 0.0689 | 0.00610 | 0.0455 | 0.126 | 0.178 |
| Reduced Form | 0.180** | 0.359*** | 1.025*** | 1.242*** | 0.286** | 0.215** |
| | (0.0794) | (0.0514) | (0.216) | (0.308) | (0.111) | (0.0928) |
| | | | | | | |
| Parent Vaccinated **w. Heckman correction** | 1.201** | 2.186*** | 6.707*** | 9.246** | 1.508** | 1.263** |
| | (0.400) | (0.652) | (1.517) | (2.780) | (0.599) | (0.504) |
| Observations | 109,112 | 109,112 | 29,748 | 29,748 | 90,294 | 90,294 |
| | | | | | | |
| **(B) Generation 3** | | | | | | |
| Grandparent Vaccinated | 1.236*** | 1.057** | 4.503*** | 4.262** | -1.0278 | -0.715 |
| | (0.361) | (0.497) | (0.916) | (1.886) | (0.836) | (0.445) |
| R sq | 0.116 | 0.187 | 0.00830 | 0.0316 | 0.0831 | 0.0846 |
| Reduced Form | 0.190*** | 0.172*** | 0.675*** | 0.661** | -0.192 | -0.140 |
| | (0.0578) | (0.0481) | (0.0199) | (0.300) | (0.171) | (0.159) |
| | | | | | | |
| Grandparent Vaccinated **w. Heckman correction** | 1.308*** | 1.124*** | 4.089*** | 4.298** | 0.927 | -0.128 |
| | (0.295) | (0.281) | (0.704) | (1.635) | (0.524) | (0.205) |
| Individuals | 116,544 | 116,544 | 40,324 | 40,324 | 70,920 | 70,920 |
| Parish of birth fixed effects | Yes | Yes | Yes | Yes | Yes | Yes |
| Year of birth fixed effects | Yes | Yes | Yes | Yes | Yes | Yes |
| Region of birth x Year of birth fixed effects | Yes | Yes | Yes | Yes | Yes | Yes |
| Families' *X*s x Year of birth fixed effects | No | Yes | No | Yes | No | Yes |
| Parish of birth *X*s x Year of birth fixed effects | No | Yes | No | Yes | No | Yes |

*Note:* Observations are *stacked* individuals (54,506 *unique* individuals for generation 2 and 58,272 *unique* individuals for generation 3 in total). First-stage regression is the same as in Table 3. The controls included are indicated in the table by Yes and No. "Families' Xs" include child characteristics at birth: sex, paternal occupational score, maternal occupational score, paternal literacy, maternal literacy, proportion of non-surviving children in the family, maternal marital status, the presence of siblings deceased due to external or unknown causes. "Parish Xs" include time-varying parish of birth characteristics: the number of midwives, the number of priests, smallpox death rate, university students per capita, price of rye, and the share of urban population. Standard errors are clustered at the (parental) parish-of-birth level. *** $p<0.001$, ** $p<0.01$, * $p<0.05$

Table 5 – Direct and mediated effects of smallpox vaccination of Generation 1 on lifetime outcomes of Generation 2 and 3: SSIV estimates with $C_{p(t-1)}$ x $C_t$ as an instrument for Generation 1 (2SRI)

| | Years lived at birth | | | | Disability-free years lived at birth | | | | Occupational score, max in ages 20-100 | | | |
|---|---|---|---|---|---|---|---|---|---|---|---|---|
| | *Generation 2(3) Mediator:* | | | | *Generation 2(3) Mediator:* | | | | *Generation 2(3) Mediator:* | | | |
| | Vaccinated in childhood | Parental occupation at birth | Midwife-assisted birth | Epigenetic factors | Vaccinated in childhood | Parental occupation at birth | Midwife-assisted birth | Epigenetic factors | Vaccinated in childhood | Parental occupation at birth | Midwife-assisted birth | Epigenetic factors |
| | (1) | (2) | (3) | (4) | (5) | (6) | (7) | (8) | (9) | (10) | (11) | (12) |
| **(A) Generation 2** | | | | | | | | | | | | |
| Parent Vaccinated (natural direct effect) | 0.772** | 1.743*** | 1.994*** | 1.411*** | 4.097*** | 6.854*** | 6.571*** | 6.896*** | 1.752*** | 2.161*** | 2.291*** | 1.610** |
| | (0.303) | (0.316) | (0.338) | (0.426) | (1.469) | (1.517) | (1.139) | (1.548) | (0.381) | (0.595) | (0.594) | (0.602) |
| Mediated Effect (natural indirect effect) | 1.042*** | -0.00587* | -0.00788*** | 0.261** | 2.749*** | -0.00266 | -0.00585* | 0.175 | 0.0636** | 0.0720** | 0.00279 | -0.101 |
| | (0.0881) | (0.00342) | (0.00171) | (0.099) | (0.405) | (0.00993) | (0.00328) | (0.226) | (0.0255) | (0.0289) | (0.00346) | (0.066) |
| Observations | 109,112 | 109,112 | 109,112 | 109,112 | 29,748 | 29,748 | 29,748 | 29,748 | 90,294 | 90,294 | 90,294 | 90,294 |
| | | | | | | | | | | | | |
| **(B) Generation 3** | | | | | | | | | | | | |
| Grandparent Vaccinated (natural direct effect) | 0.691 | 1.469*** | 1.307*** | 1.334*** | 2.690** | 5.161*** | 5.185*** | 4.510*** | 0.0819 | 0.0334 | 0.0925 | 0.358 |
| | (0.426) | (0.440) | (0.441) | (0.363) | (1.0781) | (1.124) | (1.126) | (0.919) | (0.332) | (0.331) | (0.332) | (1.056) |
| Mediated Effect (natural indirect effect) | 0.838*** | -0.00684 | -0.00209 | -0.014 | 2.504*** | 0.00561 | -0.00501 | 0.0244 | 0.0308** | 0.0866*** | 0.0161 | 0.00739 |
| | (0.124) | (0.00535) | (0.0139) | (0.362) | (0.363) | (0.0114) | (0.0530) | (0.0663) | (0.0148) | (0.0309) | (0.0137) | (0.0611) |
| Individuals | 116,544 | 116,544 | 116,544 | 116,544 | 40,324 | 40,324 | 40,324 | 40,324 | 70,920 | 70,920 | 70,920 | 70,920 |
| Parish of birth fixed effects | Yes | Yes | Yes | Yes | Yes | Yes | Yes | Yes | Yes | Yes | Yes | Yes |
| Year of birth fixed effects | Yes | Yes | Yes | Yes | Yes | Yes | Yes | Yes | Yes | Yes | Yes | Yes |
| Region of birth x Year of birth fixed effects | Yes | Yes | Yes | Yes | Yes | Yes | Yes | Yes | Yes | Yes | Yes | Yes |

*Note:* Observations are *stacked* individuals (54,506 *unique* individuals for generation 2 and 58,272 *unique* individuals for generation 3 in total). First-stage regression is the same as in Table 3. The effects are estimated for each mediator separately. Bootstrapped standard errors are clustered at the (parental) parish-of-birth level.

*** p<0.001, ** p<0.01, * p<0.05

# Online Appendices

## to the paper “Multigenerational Effects of Smallpox Vaccination”

by Volha Lazuka and Peter Sandholt Jensen

## Appendix A – Selection into Treatment and Data Accuracy

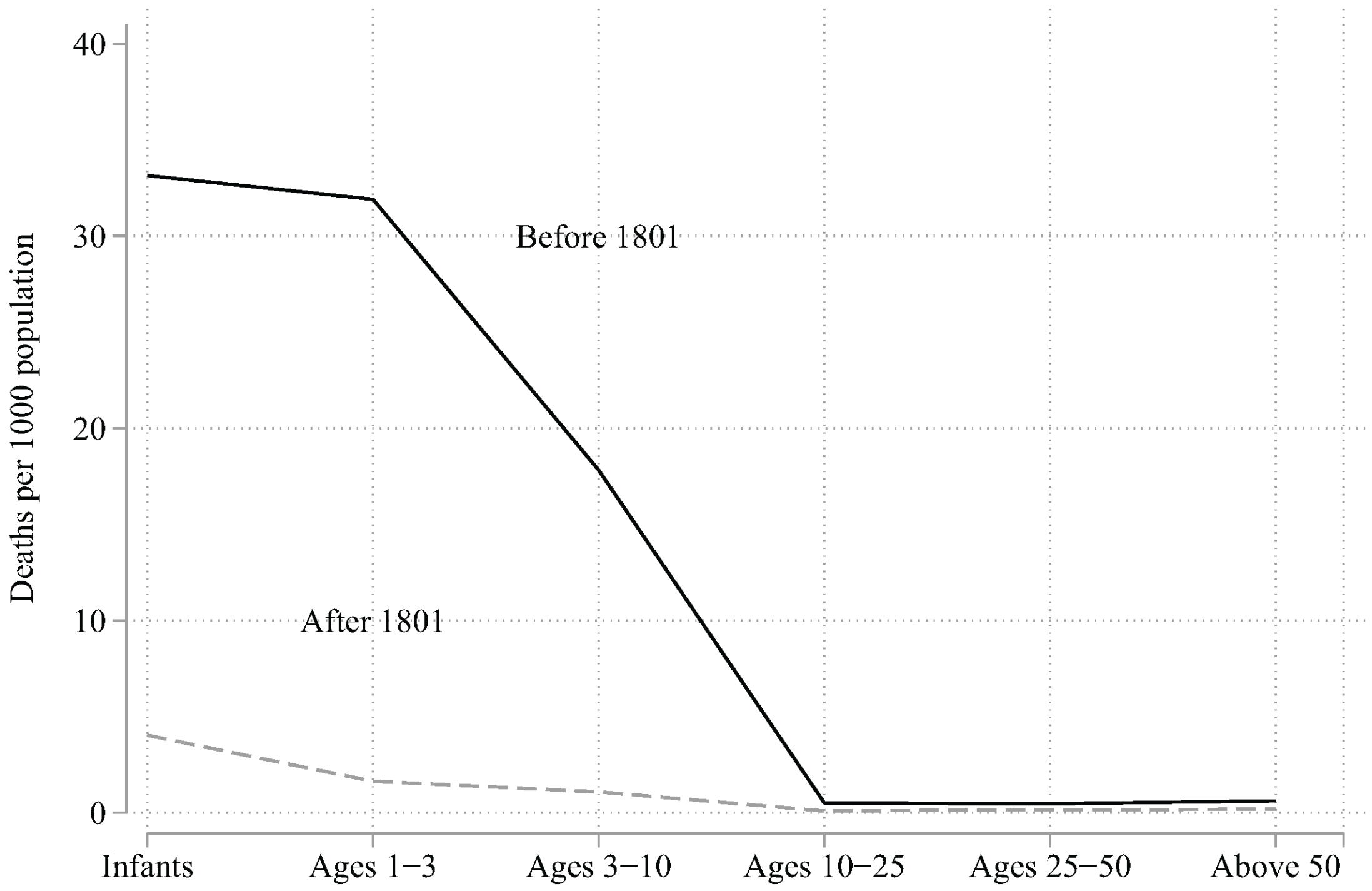

Figure A1 – Age pattern of smallpox mortality before and after the introduction of vaccination, 1790–1820.

*Source*: own calculations based on the estimation sample.

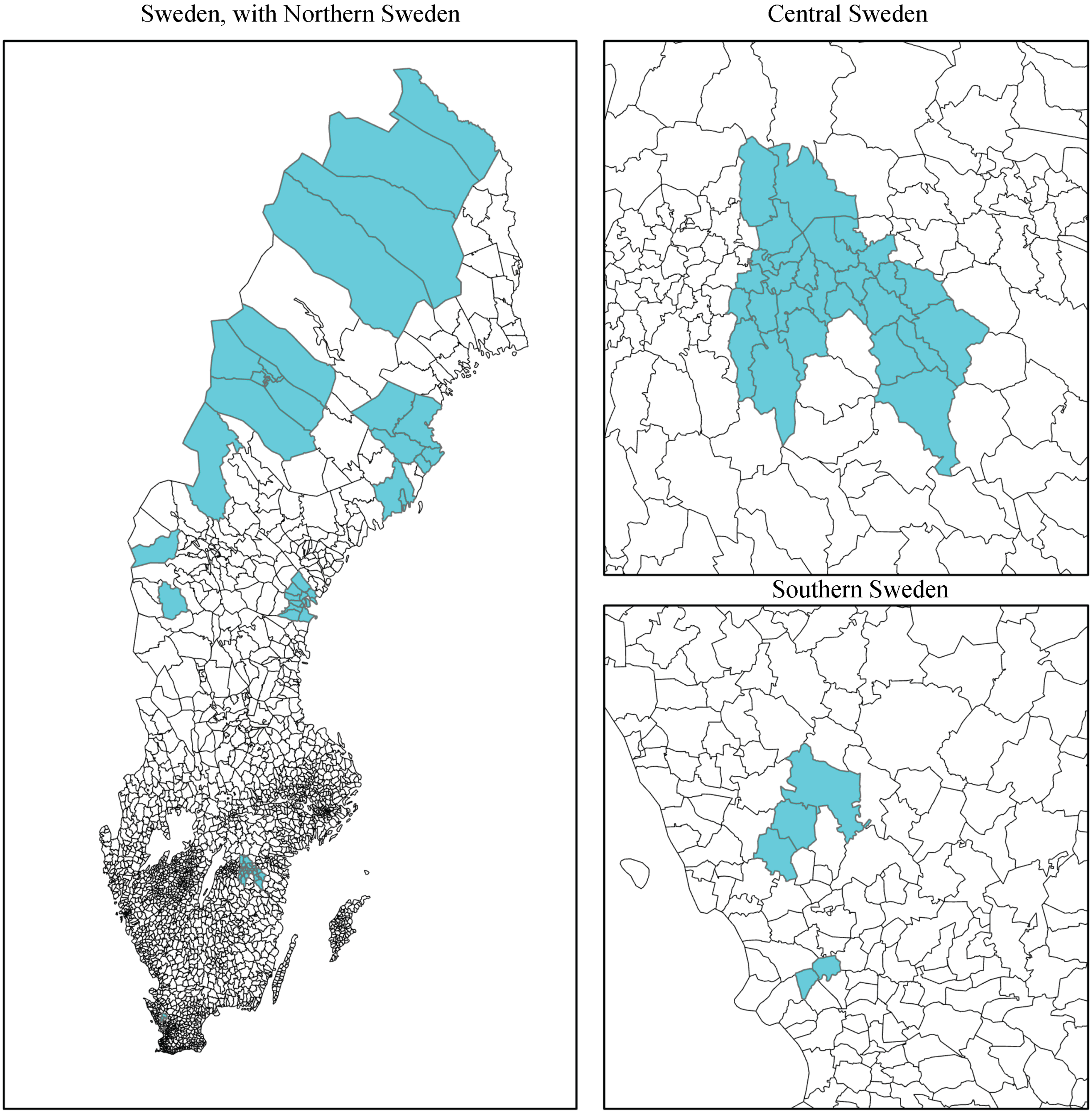

Figure A2 – Parishes under analysis, a snapshot of Sweden in 1820

*Source*: Based on the estimation sample and administrative boundaries from Riksarkivet (2016).

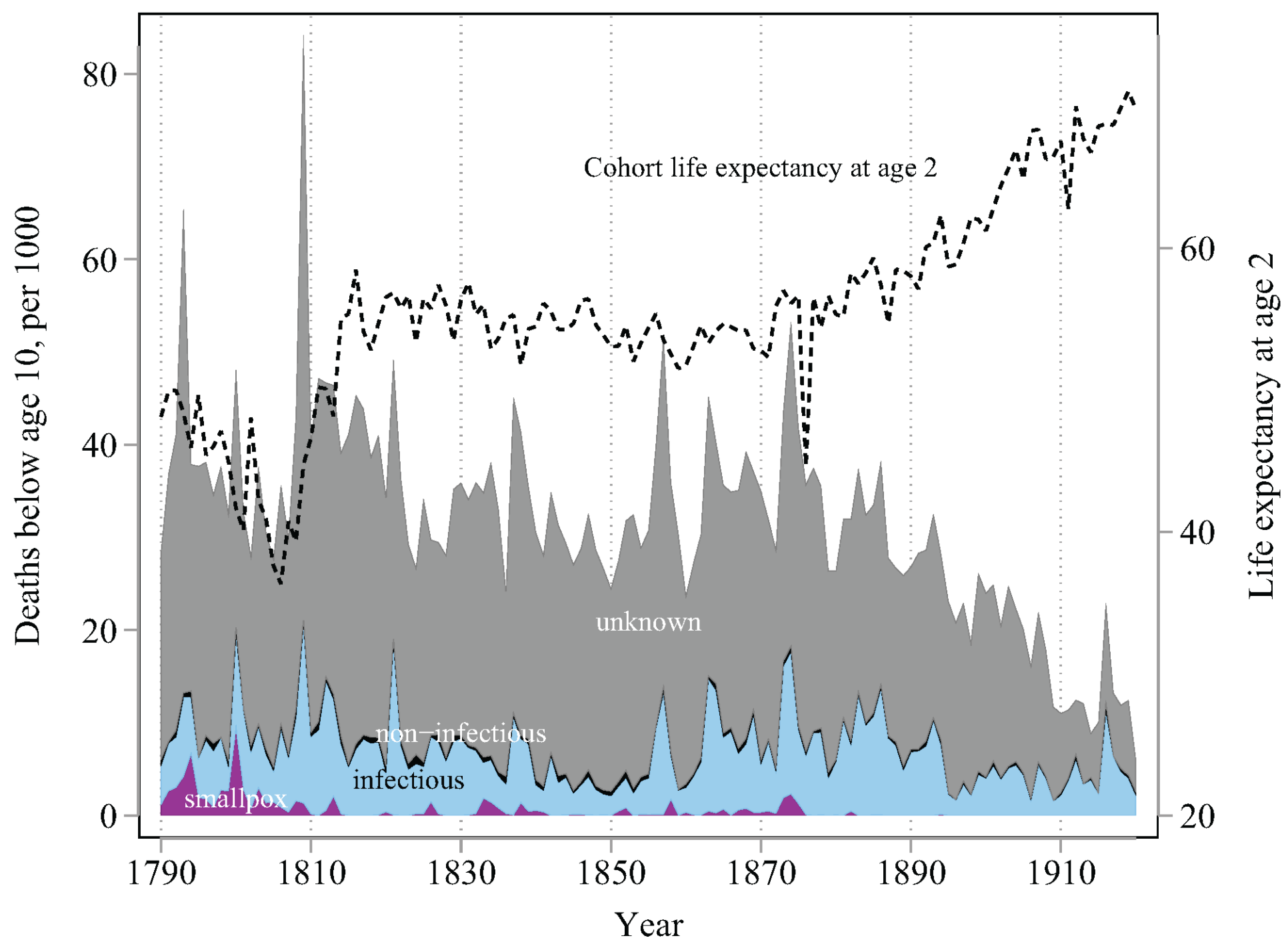

Figure A3 – Mortality rate below age 10 by cause and cohort life expectancy at age 2, 1790–1920

*Source*: own calculations based on the estimation sample.

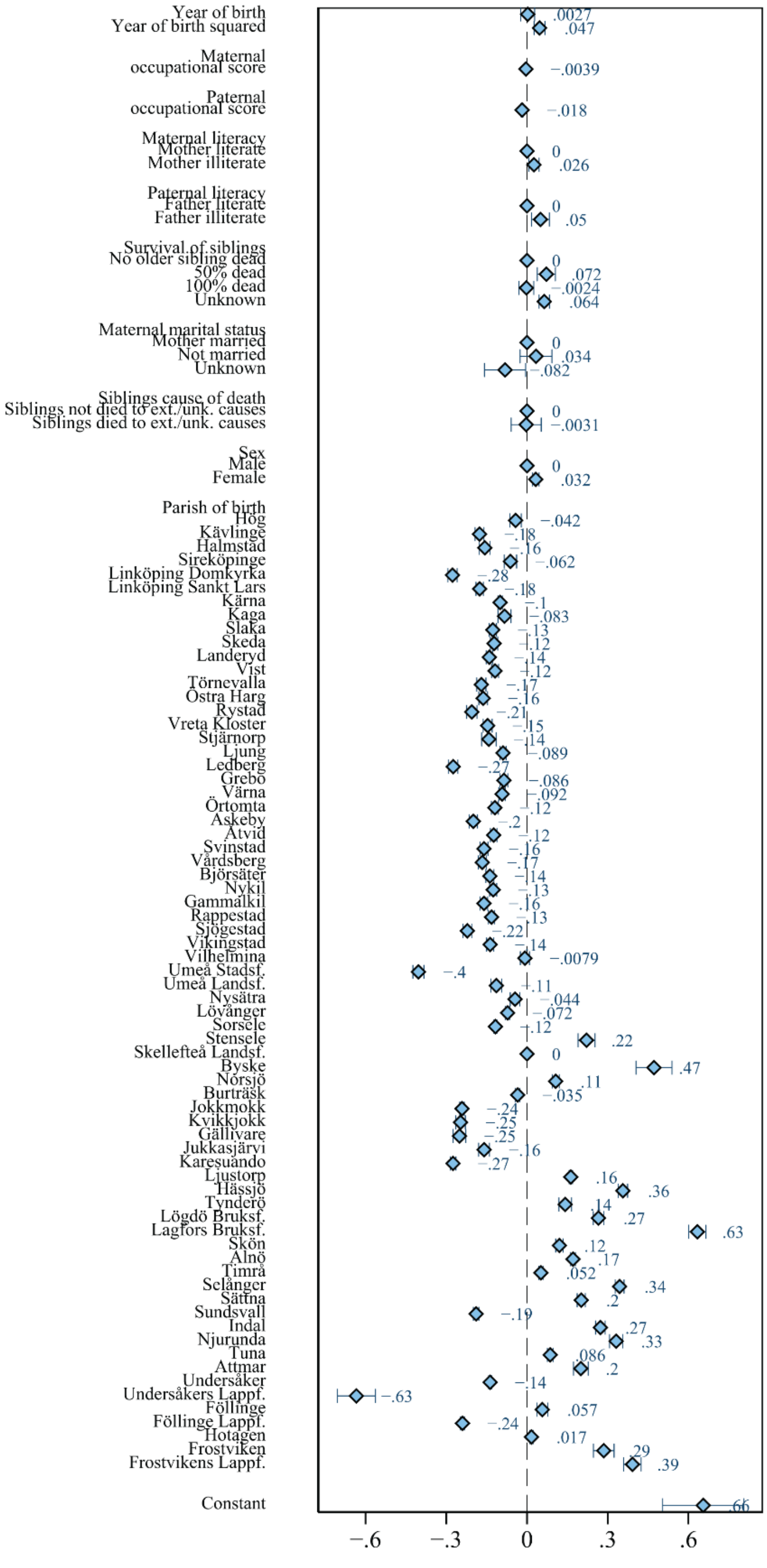

Figure A4 – Differences in the observable characteristics between individuals belonging to Generation 1, with children and grand-children (Generations 2 and 3) observed and not observed in the sample.

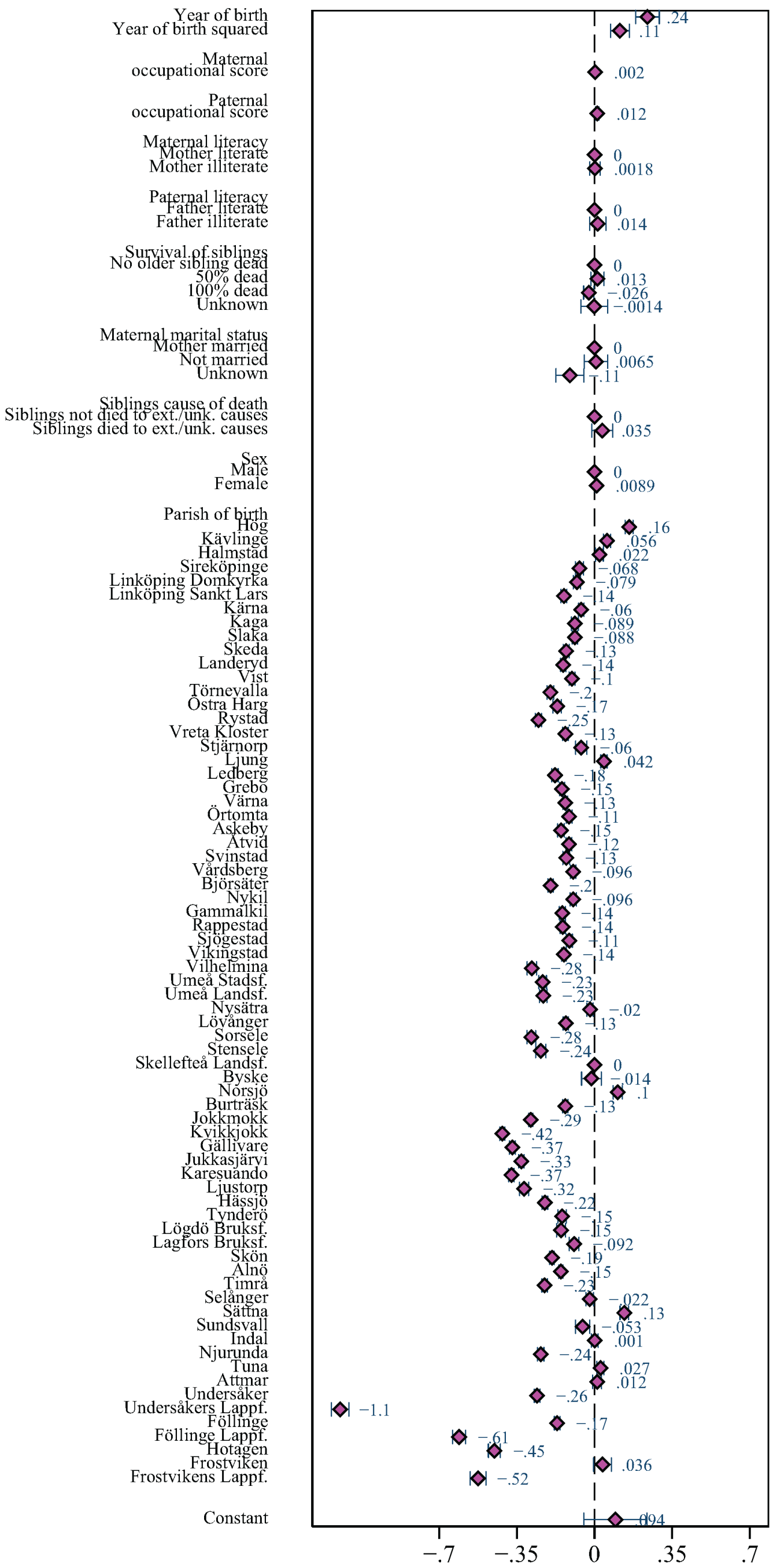

Figure A5 – Selection into vaccination status by the age of 2 for Generation 1

*Note*: The estimates are obtained from an OLS multivariate regression with a vaccination by age 2 as an outcome, cohorts born in 1790-1820. Point estimates and 95% confidence intervals. Robust standard errors are clustered at the parish of birth. Continuous variables (year of birth, year of birth squared, maternal and paternal occupational scores) are divided by their standard deviation.

Table A1 – Check for the data inaccuracy – on the whole sample

| | Vaccinated by age 2 |
|---|---|
| Spring | ref |
| Summer | 0.007* |
| | (0.004) |
| Autumn | 0.007 |
| | (0.004) |
| Winter | 0.007* |
| | (0.004) |
| Constant | 0.130*** |
| | (0.005) |
| | |
| Individuals | 43,450 |
| R-squared | 0.670 |

Note: OLS regression estimates with parish and year of birth fixed effects for the first generation. Robust standard errors in parentheses

*** p<0.01, ** p<0.05, * p<0.1

# Appendix B – Additional Results with Mother Fixed Effects.

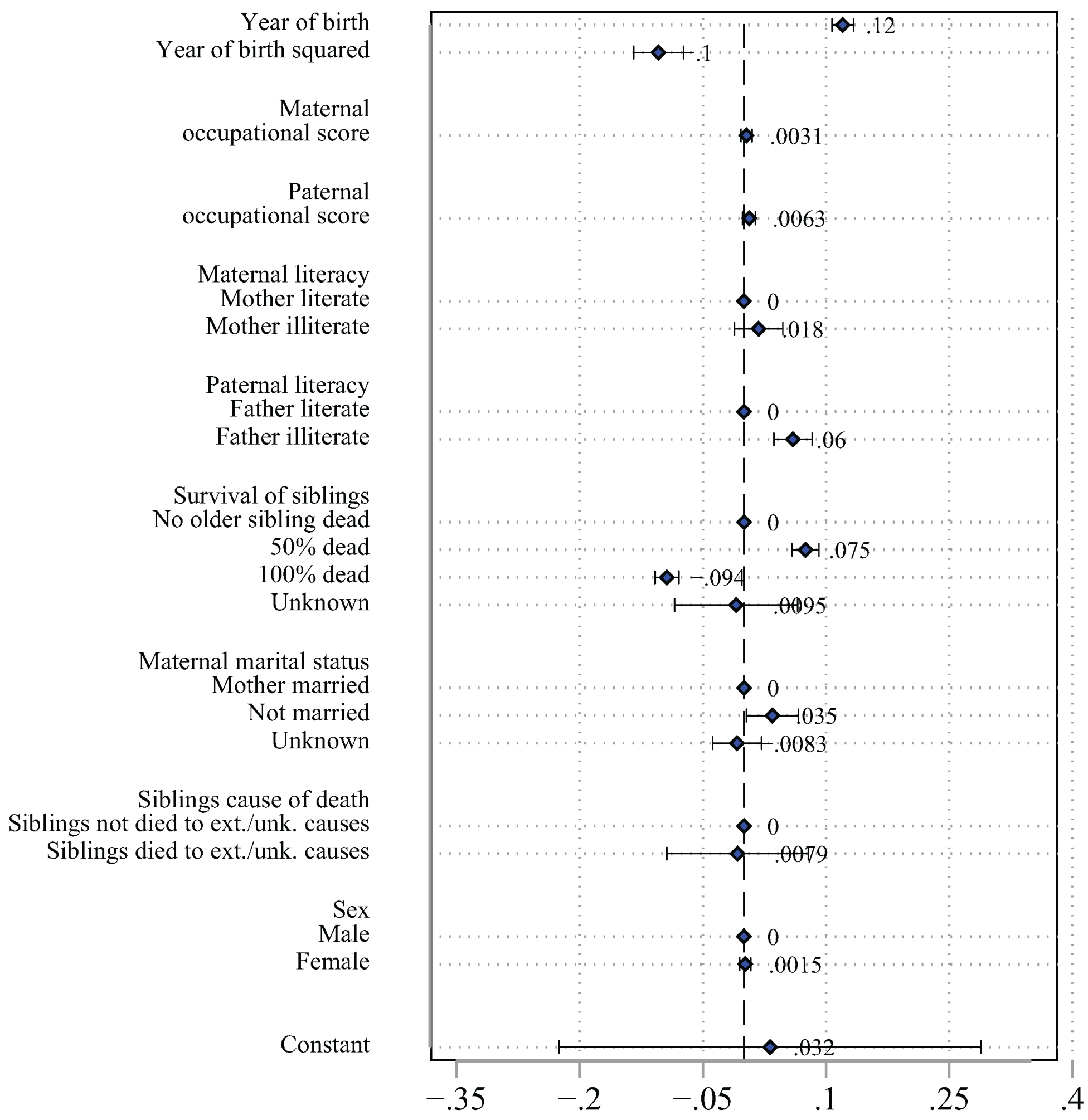

Figure B1 – Differences in the share of mothers with varying vaccination status of their children across children's observable characteristics (versus mothers with unvarying status).

*Note*: 14.5% of mothers have both vaccinated and not-vaccinated children.

Table B1 – OLS estimates of the effect of smallpox vaccination on lifetime outcomes of Generation 1: Controlling-for-observables

| | Remaining years lived at age 2 | | Disability-free years lived at age 2 | | Has good literacy, after age 15 | | Occupational score, max in ages 15-100 | |
|---|---|---|---|---|---|---|---|---|
| | (1) | (2) | (3) | (4) | (5) | (6) | (7) | (8) |
| **(I) FULL SAMPLE** | | | | | | | | |
| **Panel A: Baseline estimates** | | | | | | | | |
| Vaccinated | 11.697**** | 17.423**** | 11.147*** | 17.429*** | 0.188*** | 0.108*** | 3.802*** | 4.481*** |
| | (0.198) | (0.341) | (0.197) | (0.347) | (0.00432) | (0.00438) | (0.169) | (0.177) |
| Rsq | 0.074 | 0.170 | 0.070 | 0.168 | 0.062 | 0.148 | 0.0016 | 0.0211 |
| **Panel B: Heckman-corrected estimates** | | | | | | | | |
| Vaccinated | 9.266*** | 17.224*** | 8.348*** | 17.514*** | 0.0381*** | 0.0390*** | 4.961*** | 4.983*** |
| | (0.214) | (0.339) | (0.211) | (0.344) | (0.00772) | (0.0071) | (0.345) | (0.344) |
| Rsq | 0.091 | 0.182 | 0.096 | 0.183 | 0.221 | 0.221 | 0.130 | 0.130 |
| Observations | 43,450 | 43,450 | 42,021 | 42,021 | 28,614 | 28,614 | 30,806 | 30,806 |
| **(II) MOTHER FE SAMPLE** | | | | | | | | |
| **Panel A: Baseline estimates** | | | | | | | | |
| Vaccinated | 15.042**** | 17.267*** | 14.885*** | 16.882*** | 0.0906*** | 0.0604*** | 2.449*** | 3.547*** |
| | (0.491) | (0.654) | (0.481) | (0.655) | (0.00754) | (0.011) | (0.413) | (0.580) |
| Rsq | 0.129 | 0.223 | 0.132 | 0.210 | 0.017 | 0.242 | 0.0037 | 0.143 |
| **Panel B: Heckman-corrected estimates** | | | | | | | | |
| Vaccinated | 15.085*** | 17.270*** | 14.763*** | 16.882*** | 0.0914*** | 0.0553*** | 2.581*** | 3.604*** |
| | (0.492) | (0.654) | (0.482) | (0.655) | (0.00758) | (0.0106) | (0.471) | (0.576) |
| Rsq | 0.129 | 0.223 | 0.133 | 0.210 | 0.017 | 0.293 | 0.0036 | 0.144 |
| Observations | 6,313 | 6,313 | 6,258 | 6,258 | 28,614 | 28,614 | 2.668 | 2.668 |
| Parish of birth FEs | No | Yes | No | Yes | No | Yes | No | Yes |
| Year of birth FEs | No | Yes | No | Yes | No | Yes | No | Yes |
| Region of birth x Year of birth FEs | No | Yes | No | Yes | No | Yes | No | Yes |
| Families' Xs x Year of birth FEs | No | Yes | No | Yes | No | Yes | No | Yes |
| Parish of birth Xs x Year of birth FEs | No | Yes | No | Yes | No | Yes | No | Yes |

*Note:* Observations are individuals. Panel A reports the estimates based on Equation 1. Panel B reports the estimates with a Heckman correction (see section 3.4 for the procedure). The controls included are indicated in the table by Yes and No. "Families' Xs" include child characteristics at birth: sex, paternal occupational score, maternal occupational score, paternal literacy, maternal literacy, proportion of non-surviving children in the family, maternal marital status, the presence of siblings deceased due to external or unknown causes. "Parish Xs" include time-varying parish of birth characteristics: the number of midwives, the number of priests, smallpox death rate, university students per capita, price of rye, and the share of urban population.

*** p<0.001, ** p<0.01, * p<0.05

# Appendix C – Additional Results with SSIV.

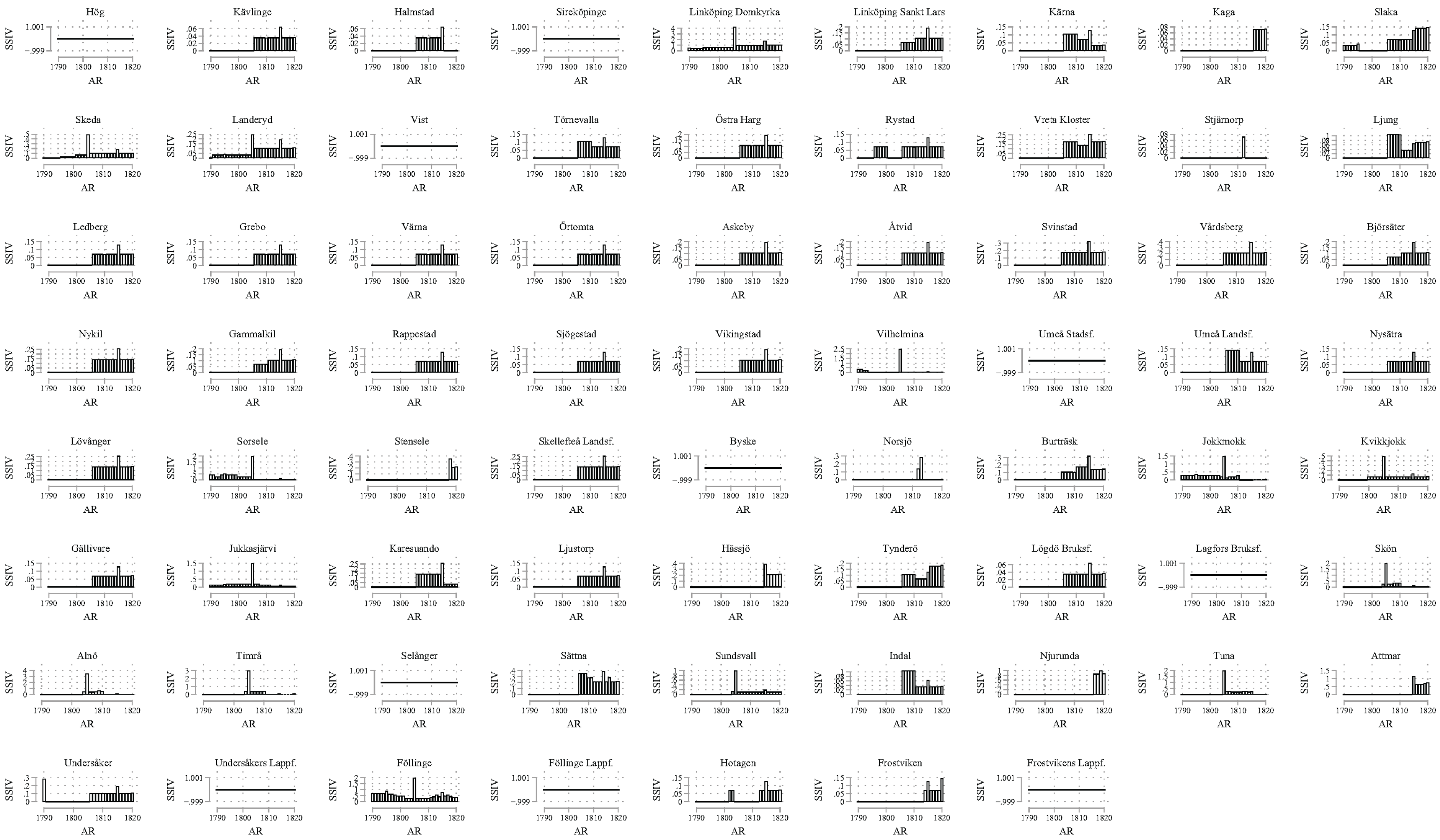

Figure C1 – The values of the SSIV ($C_{p(t-1)}$ x $C_t$) across parishes and cohorts.

*Source*: own calculations from the estimation sample.

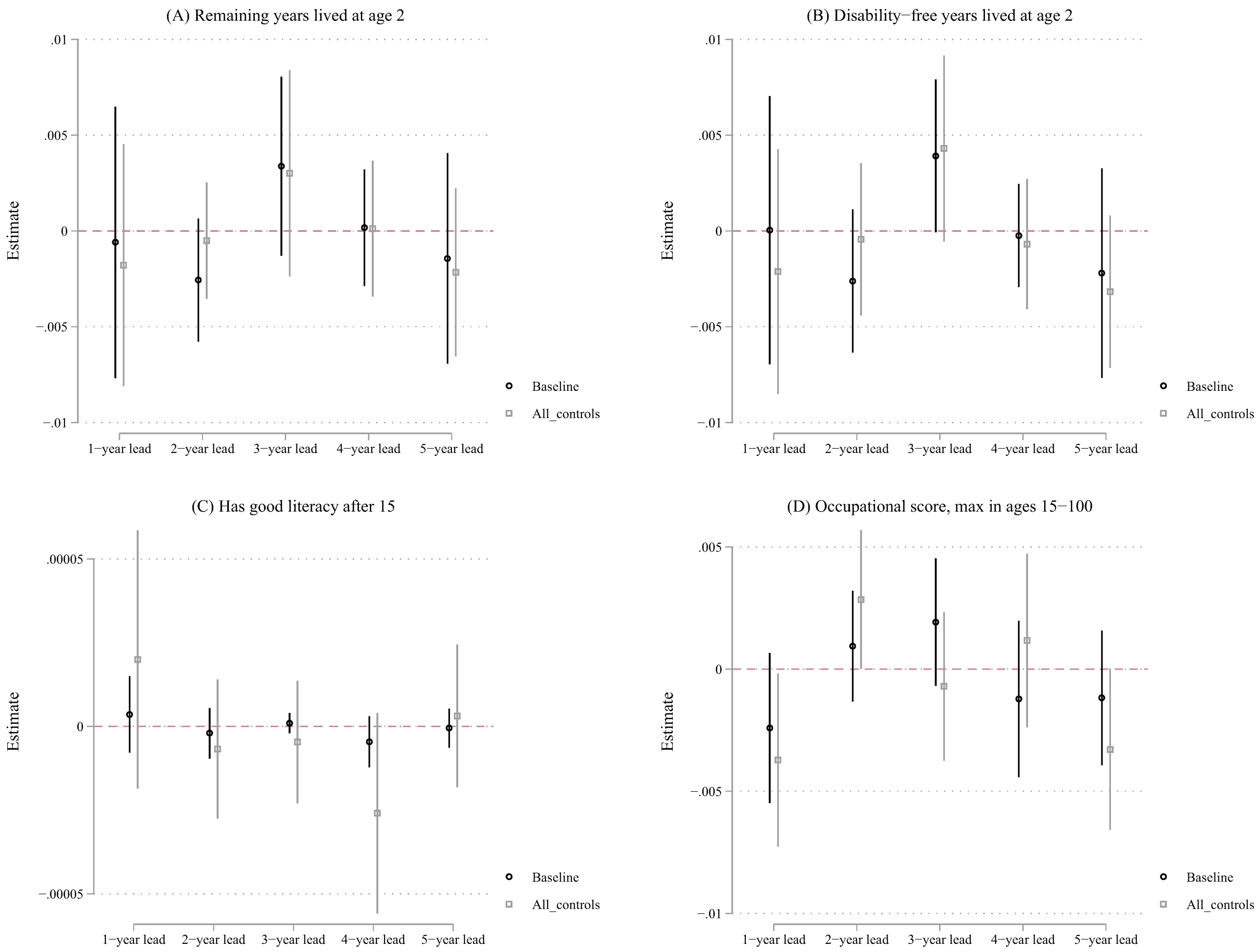

Figure C2 – 2SLS estimates of the leads of the interacted IV for all outcomes.

*Note:* 2SLS point estimates and 95% confidence intervals. We estimate the models with all 30 leads (the year of 1790 is the reference) but report the first five estimates, which is conventional in the difference-in-differences literature (Roth et al. 2023). "Baseline" controls include year of birth fixed effects, arish of birth fixed effects, and county of birth-by-year of birth fixed effects. "All controls" additionaly include interactions between year of birth and child characteristics at birth (sex, paternal occupational score, maternal occupational score, paternal literacy, maternal literacy, proportion of non-surviving children in the family, maternal marital status, the presence of siblings deceased due to external or unknown causes) and interactions between year of birth and parish of birth characteristics (the number of midwives, the number of priests, smallpox death rate, university students per capita, price of rye, and the share of urban population). Standard errors are clustered at the parish-of-birth level.

## Appendix D – The Influence of Overlapping Interventions

Table D1 – The interaction effects of vaccination with cointerventions, Generation 1: Reduced-form estimates with $C_{p(t-1)}$ x $C_{rt}$ as an instrument

| | Remaining years lived, at age 2 | | | Disability-free years lived, at age 2 | | | Occupational score, max in ages 15-100 | | |
|---|---|---|---|---|---|---|---|---|---|
| | Epidemics | Midwives | Potatoes | Epidemics | Midwives | Potatoes | Epidemics | Midwives | Potatoes |
| | (1) | (2) | (3) | (4) | (5) | (6) | (10) | (11) | (12) |
| $C_{p(t-1)}$ x $C_{rt}$ | 1.698*** | 1.567*** | 1.687*** | 1.695*** | 1.465** | 1.603*** | 3.116*** | 3.419*** | 4.164*** |
| | (0.600) | (0.602) | (0.521) | (0.573) | (0.621) | (0.402) | (0.777) | (0.747) | (1.877) |
| $C_{p(t-1)}$ x $C_{rt}$ X Cointervention | 1.021** | -0.0142 | -0.121 | 0.951** | 0.0935 | -0.123 | 2.329*** | -0.0567 | 0.367 |
| | (0.424) | (0.0515) | (0.122) | (0.432) | (0.725) | (0.268) | (0.809) | (0.0465) | (0.297) |
| R sq | 0.112 | 0.113 | 0.113 | 0.103 | 0.106 | 0.103 | 0.158 | 0.158 | 0.158 |
| Observations | 32,120 | 32,120 | 32,120 | 30,930 | 30,930 | 30,930 | 22,823 | 22,823 | 22,823 |
| Parish of birth FEs | Yes | Yes | Yes | Yes | Yes | Yes | Yes | Yes | Yes |
| Year of birth FEs | Yes | Yes | Yes | Yes | Yes | Yes | Yes | Yes | Yes |
| Region of birth x Year of birth FEs | Yes | Yes | Yes | Yes | Yes | Yes | Yes | Yes | Yes |

*Note:* "Epidemics": The cointervention variable is smallpox mortality among children under age 10. "Midwives": The cointervention variable is the ratio of midwives to the population. "Potatoes": The cointervention variable is the quantity of potato seeds per square kilometer. All variables are divided by their means to facilitate interpretation. Standard errors are clustered at the parish-of-birth level.

*** p<0.001, ** p<0.01, * p<0.05

# References (in Appendices)